\documentclass[aps,notitlepage,superscriptaddress,nofootinbib,twocolumn,floatfix]{revtex4-1}
\usepackage[utf8]{inputenc}

\usepackage[table]{xcolor}

\usepackage{multirow}

\usepackage{hyperref}
\usepackage{graphicx}
\usepackage{mathtools}
\usepackage{amsmath,amssymb,amsfonts,amsthm,latexsym,stmaryrd,physics,mathrsfs}
\allowdisplaybreaks[4]
\usepackage{marginnote}
\usepackage{color}
\usepackage{bm}
\usepackage{float}
\usepackage{nicefrac}
\usepackage{microtype} 
\usepackage{booktabs}
\usepackage{verbatim}
\usepackage{setspace}
\usepackage[T1]{fontenc}
\usepackage[utf8]{inputenc}
\usepackage{stfloats}
\usepackage{diagbox}
\usepackage{dsfont}

\hypersetup{
	colorlinks=true
,urlcolor=blue
,anchorcolor=blue
,citecolor=blue
,filecolor=blue
,linkcolor=blue
,menucolor=blue
,pagecolor=blue
,linktocpage=true
,pdfproducer=medialab
}

\def\beq{\begin{equation}}
\def\eeq{\end{equation}}
\newcommand{\bea}{\begin{eqnarray}}
\newcommand{\eea}{\end{eqnarray}}
\def\bi{\begin{itemize}}
\def\ei{\end{itemize}}
\def\ba{\begin{array}}
\def\ea{\end{array}}
\def\bfig{\begin{figure}}
\def\efig{\end{figure}}

\def\C{\mathbb{C}}
\def\R{\mathbb{R}}
\def\Z{\mathbb{Z}}

\def\sgn{\text{sgn}}

\newcommand{\Slc}{\mathrm{SL}(2,\mathbb{C})}

\newcommand{\Su}{\mathrm{SU}(2)}

\def\be{\begin{eqnarray}}
\def\ee{\end{eqnarray}}

\newcommand{\ck}{\mathcal K}

\newcommand{\calp}{\mathcal P}

\newcommand{\cs}{\mathcal S}

\newcommand{\cz}{\mathcal Z}

\newcommand{\fa}{\mathfrak{a}}

\newcommand{\fl}{\mathfrak{l}}

\renewcommand{\a}{\alpha}
\renewcommand{\b}{\beta}
\newcommand{\g}{\gamma}

\renewcommand{\l}{\lambda}

\renewcommand{\O}{\Omega}

\newcommand{\rmd}{\mathrm d}

\newcommand{\lt}{\left}
\newcommand{\rt}{\right}

\newcommand{\act}{\rhd}

\newcommand{\re}{\mathrm{Re}}

\begin{document}
\title{\boldmath Cosmological Dynamics from Covariant Loop Quantum Gravity with Scalar Matter}

\author{Muxin Han}
\email{hanm@fau.edu}
\affiliation{Department of Physics, Florida Atlantic University, 777 Glades Road, Boca Raton, FL 33431-0991, USA}
\affiliation{Department Physik, Institut f\"ur Quantengravitation, Theoretische Physik III, Friedrich-Alexander Universit\"at Erlangen-N\"urnberg, Staudtstr. 7/B2, 91058 Erlangen, Germany}

\author{Hongguang Liu} 
\email{hongguang.liu@gravity.fau.de}
\affiliation{Department Physik, Institut f\"ur Quantengravitation, Theoretische Physik III, Friedrich-Alexander Universit\"at Erlangen-N\"urnberg, Staudtstr. 7/B2, 91058 Erlangen, Germany}

\author{Dongxue Qu}
\email{dqu@perimeterinstitute.ca}
\affiliation{Perimeter Institute for Theoretical Physics, 31 Caroline St N, N2L\,2Y5 Waterloo, ON, Canada}

\author{Francesca Vidotto}
\email{fvidotto@uwo.ca}
\affiliation{Department of Physics and Astronomy, Department of Philosophy and Rotman Institute, Western University, 1151 Richmond Street, N6A\,3K7 London, ON, Canada}

\author{Cong Zhang}
\email{zhang.cong@mail.bnu.edu.cn}
\affiliation{Department Physik, Institut f\"ur Quantengravitation, Theoretische Physik III, Friedrich-Alexander Universit\"at Erlangen-N\"urnberg, Staudtstr. 7/B2, 91058 Erlangen, Germany}

\begin{abstract}
We study homogenous and isotropic quantum cosmology using the spinfoam formalism of Loop Quantum Gravity (LQG). We define a coupling of a scalar field to the 4-dimensional Lorentzian Engle-Pereira-Rovelli-Livine (EPRL) spinfoam model. We employ the numerical method of complex critical points to investigate the model on two different simplicial complexes: the triangulations of a single hypercube and two connected hypercubes. We find nontrivial implications for the effective cosmological dynamics. In the single-hypercube model, the numerical results suggest an effective Friedmann equation with a scalar density that contains higher-order derivatives and a scalar potential. The scalar potential plays a role similar to a positive cosmological constant and drives an accelerated expansion of the universe. The double-hypercubes model resembles a symmetric cosmic bounce, and a similar effective Friedmann equation emerges with higher-order derivative terms in the effective scalar density, whereas the scalar potential becomes negligible.
\end{abstract}

\maketitle





\section{Introduction}

Understanding the early dynamics of the universe has been a central quest in theoretical physics. 
A quantum theory of gravity is needed for this, and Loop Quantum Gravity (LQG) stands out as a promising approach in this endeavor. The study of the cosmological dynamics within the framework of LQG offers a novel perspective on the quantum nature of the gravitational field and its interaction with matter fields. 
The main approach for studying cosmology in LQG, Loop Quantum Cosmology (LQC), is based on the canonical formulation of the theory. For a comprehensive review we refer the reader, for example, to \cite{Agullo:2016tjh}. This paper focuses instead on the dynamics of LQG in its covariant form, the \emph{spinfoam} formalism \cite{Perez:2012wv,Rovelli:2014ssa}. Our approach develops the understanding of quantum cosmology from the full covariant LQG theory, in contrasts to LQC, which is based on the symmetry reduced models.

The application of the covariant LQG dynamics in cosmology was introduced in \cite{Bianchi:2010zs, Vidotto:2011qa}.
The idea of this approach is to explore the approximation by studying the non-perturbative behavior of a finite number of degrees of freedom (e.g. by considering simple boundary graph states and by truncating the spinfoam expansion to the first terms \cite{Rovelli2008d,Borja:2011di,Vidotto:2015bza}). 
The covariant dynamics remains well defined in the deep quantum regime, making possible to study the properties of the quantum state of the universe in its ealiest phases. Recent developments in this directions have aimed at predicting primordial quantum fluctuations \cite{Gozzini:2019nbo,Frisoni:2022urv,Frisoni:2023lvb}. The main obstacle in this research program has been the difficulty of the computations, for which the development of numerical methods has been essential.



Recently, there has been a considerable growth of interest in numerical methods for covariant LQG. 
Among various numerical approaches in spinfoams (see \cite{Dona:2022yyn,Asante:2022dnj} for recent reviews), the main numerical tool we employ in this paper is the method of \emph{complex critical points}, a numerical approach for investigating the oscillatory integral representation of the Lorentzian spinfoam amplitude. This approach closely relates to the stationary phase approximation, a key tool for studying quantum theory by the perturbative expansion. One of the interesting aspects in earlier numerical results \cite{Han:2021kll,Han:2023cen} is extracting properties of effective theory from the spinfoam amplitude in the large-$j$ regime. In this paper, we apply this method to the 4d Lorentzian EPRL spinfoam amplitude with the initial/final condition corresponding to the homogeneous and isotropic cosmology. The purpose of this paper is to investigate the effective dynamics of cosmology obtained from the large-$j$ spinfoam amplitude. We build on some earlier results from canonical LQG,
e.g., \cite{Han:2019vpw,Zhang:2020mld,Han:2020iwk, Zhang:2021qul, Han:2021cwb,Dapor:2017rwv}.

Matter plays a crucial role in the cosmological dynamics; therefore, it is necessary to couple matter to spinfoams in the present analysis. Although there were early investigations of coupling spinfoams to scalars, fermions, and gauge fields in 4d, e.g., \cite{Bianchi:2010bn,Han:2011as,Ali:2022vhn}, the physical applications have not been extensively explored so far. In this paper, we consider a scalar field coupled to spinfoams. The numerical results demonstrate a nontrivial physical impact from the presence of the scalar field on the effective cosmological dynamics. 


In this paper, we study the spinfoam amplitude coupled with the scalar field on two types of 4d simplicial complexes: the single-hypercube complex and the double-hypercube complex. These complexes are respectively the simplicial triangulations of a single hypercube and two connected hypercubes. Both complexes are periodic along the spatial direction and are triangulations of $\mathbb{T}^3\times [0,1]$. In relation to spatially flat cosmology, the triangulated 3d cube (triangulating $\mathbb{T}^3$) in the complexes at a fixed time is understood as an elemetary cell in space. The periodicity models the spatial homogeneity, and requiring the cube to be equilateral is the discrete analog of isotropy.

In the single-hypercube model, we set the initial data $(a_i, K_i, \phi_i, \pi_i)$ satisfying the Friedmann equation on the triangulated cube at a $t=0$ slice. Here $K_{i(f)}$ is the extrinsic curvature, $a_{i(f)}$ is the edge length of cubes, and $(\phi_{i(f)}, \pi_{i(f)})$ are the canonical pair of the scalar field. We perform the numerical computation of the complex critical points of the spinfoam amplitude with a large number of final data samples $(a_f, K_f, \phi_f, \pi_f)$. The amplitude evluated at the complex critical point provides the leading-order contribution in the large-$j$ limit. If we fix $a_f,\phi_f,\pi_f$ and consider the absolute-value of the amplitude as the function of $K_f$, we compute the location, denoted by $K_{\rm crit}$, of the maximum of this function. 
When we vary $\phi_f,\pi_f$, the value of $K_{\rm crit}$ varies 
accordingly. The numerical results reveal a polynomial dependence of the form $K^2_{\rm crit} = \alpha_0 (\phi_f) + \alpha_2(\phi_f)\pi^2_f + \alpha_3(\phi_f)\pi^3_f + \alpha_4(\phi_f)\pi^4_f + O(\pi^5_f)$. This equation represents an effective constraint on the final data from the large-$j$ spinfoam amplitude. The constraint resembles a modified Friedmann equation with an effective scalar density $\rho_{\rm eff}$ as the right-hand side of the equation, containing not only $\pi^2$ but also higher derivative terms $\pi^3,\pi^4,\cdots$. The effective scalar density also includes $\alpha_0 (\phi_f)$, which acts as an effective scalar potential. The result $\alpha_0>0$ plays a role similar to an effective positive cosmological constant. Moreover,  the non-zero $\alpha_0$ relates to the result that on the final slice $K_{\rm crit} > K_i$, indicating an accelerated expansion of the universe.

The double-hypercubes model aims to provide a spinfoam analog of the (time-reversal) symmetric bounce in LQG. Intuitively, the spacelike cube shared by two hypercubes represents the instance of the bounce. The computation scheme is similar to the single-hypercube model. To create an analog to the symmetric bounce, we constrain the initial and final data by the conditions: $a_i=a_f=a, K_f=-K_i>0, \phi_f=-\phi_i>0, \pi_f=\pi_i>0$. The property that $K=\dot{a}$ is negative at the initial and positive at the final indicates contracting and expanding universes at the initial and final slices. The evolution from negative to positive values of $\dot{a}$ suggests a cosmic bounce occurring in the evolution. We again denote by $K_{\rm crit}$ the value of $K_i$ where the aboslute-value of the amplitude reaches the maximum at fixed $a_i,\phi_i,\pi_i$. The numerical results reveal that the relationship between $K^2_{\rm crit}$ and $\pi_i$ is constrained by $K^2_{\rm crit} = \alpha_2(\phi_i)\pi^2_i+\alpha_3(\phi_i)\pi^3_i+\alpha_4(\phi_i)\pi^4_i+O(\pi^5_i)$. Similar to the single-hypercube, this constraint resembles a modified Friedmann equation, and the right-hand side is proportional to the effective scalar density $\rho_{\rm eff}$, which contains higher-order derivative terms with $\pi_i^3,\pi_i^4,\cdots$. One important difference is that $\alpha_0$ is negligible for the symmetric bounce. 

Some recent analyses on the de Sitter space from the effective spinfoam model without matter coupling in \cite{Dittrich:2023rcr}, on cosmology from Regge calculus based on a Wick rotation in \cite{Jercher:2023csk}, and on cosmology from the Euclidean cuboid spinfoam model in \cite{Bahr:2017eyi}, might be related to our results.

We present our results in this article following this structure.  We discuss the Lorentzian spinfoams action for different types of triangles in Section \ref{SFAmplitude}. Section \ref{SFAmplitude} also includes a concise discussion of the spinfoam amplitude with coherent spin-network boundary state and reviews the algorithm for computing the complex critical points. In Section \ref{Complex}, we discuss the setups for the single-hypercube complex and double-hypercubes complex with periodic boundary conditions. In Section \ref{Spinfoam with Scalar Matter}, we discuss the spinfoam model coupled with the scalar field. Finally, in Section \ref{Implementation and Numerical Results}, we numerically investigate the spinfoam amplitude coupled with scalar matter on both the single-hypercube and double-hypercubes complexes.


\section{Spinfoam Amplitude}\label{SFAmplitude}

A 4-dimensional simplicial complex $\ck$ contains 4-simplices $v$, tetrahedra $e$, triangles $f$, line segments, and points. The internal and boundary triangles are denoted by $h$ and $b$ (where $f$ is either $h$ or $b$). We define the spinfoam amplitude on $\ck$ using the Hnybida-Conrady extended spinfoam model, which allows not only spacelike tetrahedra and triangles but also timelike ones on $\ck$ \cite{Conrady:2010vx,Conrady:2010kc}. The half-integer ``spins'' $j_h,j_b$, assigned to internal and boundary triangles $h$ and $b$, respectively, are related to representations of SU(2) and SU(1,1) for triangles belonging to spacelike and timelike tetrahedra. They label the quanta of triangle areas. In the large-$j$ regime, the quantum area of a spacelike triangle $f$ is given by $\mathrm{Ar}_f\simeq \g j_f$\cite{Rovelli1995,ALarea}, and $\mathrm{Ar}_f=j_f$ for timelike triangles, assuming the unit is set such that $8\pi G\hbar=1$. Here, $\g$ is the Barbero-Immirzi parameter.

The Lorentzian spinfoam amplitude on $\ck$ is given by summing over internal spins $\{j_h\}$:
\be
A(\mathcal{K})=\sum_{\{j_h\}}^{j^{\rm max}} \prod_h \mu_h(j_h) \int[\mathrm{d} X] e^{S[j_h, X;j_b,\xi_{eb}]}. \label{SFamplitude}
\ee 
In this formulation, we assume that all boundary tetrahedra are spacelike, so the boundary states of $A(\ck)$ are SU(2) coherent states $|j_b,\xi_{eb}\rangle$, where $\xi_{eb}=u_{eb}\act(1,0)^\mathrm{T}$, and $u_{eb}\in \Su$. The values of $j_b$ and $\xi_{eb}$ are determined by the area and the 3-normal of the triangle $b$ in the boundary tetrahedron $e$. $\mu_h(j_h)$ is the face amplitude, and its explicit expression does not affect the discussion in this paper. The cut-offs of the spin sums, denoted by $j^{\rm max}=\{j^{\rm max}_h\}_h$, may be implied by the triangle inequality and $j_b$ or otherwise have to be imposed by hand.  

The set of integrated variables denoted by $X$ contains some $\Slc$ group elements $g_{ve}$ and some spinor variables. The spinfoam action $S$ in (\ref{SFamplitude}) is complex and linear with respect to $j_h$ and $j_b$. The action $S=\sum_f S_f$ is a sum of face actions $S_f$ and has three types of contributions from (1) spacelike triangles in spacelike tetrahedra, (2) spacelike triangles in timelike tetrahedra, and (3) timelike triangles. Some details of the spinfoam action and integration variables are discussed below. Further details regarding the derivation of the spinfoam action $S$ can be found in \cite{Han:2013gna,Liu:2018gfc,Han:2021rjo}.

\subsection{Spinfoam action}

Different types of triangles correspond to different contributions to $S$, classified below. They are characterized by different sets of variables in $X$ and functions $F_f[X]$ that are independent of $j_f$. 
\\[.7em]
\emph{Spacelike triangles in spacelike tetrahedra:} The contribution results from the Lorentzian EPRL spinfoam model \cite{Engle:2007wy}. The corresponding integration variables in $X$ are given by
\be
\left\{ g_{ve}, z_{vf}\right\}\subset X. \label{Xdata} 
\ee 
Here, $g_{ve} \in \mathrm{SL}(2, \mathbb{C})$, and $z_{vf} \in \mathbb{CP}^1$. The spinfoam action for spacelike triangles $h,b$ in spacelike tetrahedra $e,e'$ is given by \cite{Barrett:2009mw,Han:2011re,Han:2013gna}:
\be
S_{\textit{s-s}} & =&\sum_{e^{\prime}} j_h F_{\left(e^{\prime}, h\right)}+\sum_{(e, b)} j_b F_{(e, b)}^{in/out }+\sum_{\left(e^{\prime}, b\right)} j_b F_{\left(e^{\prime}, b\right)}^{in / out },
\ee
where $e$ and $e'$ represent boundary and internal tetrahedra, respectively. The functions $F_{(e,b)}^{in/out}$ and $F_{(e',f)}$ are given by:
\be
F_{(e, b)}^{{out }} & =&2 \ln \frac{\left\langle Z_{v e b}, \xi_{e b}\right\rangle}{\left\|Z_{v e b}\right\|}+i \gamma \ln \left\|Z_{v e b}\right\|^2, \\
F_{(e, b)}^{i n} & =&2 \ln \frac{\left\langle\xi_{e b}, Z_{v^{\prime} e b}\right\rangle}{\left\|Z_{v^{\prime} e b}\right\|}-i \gamma \ln \left\|Z_{v^{\prime} e b}\right\|^2, \\
F_{\left(e^{\prime}, f\right)} & =&2 \ln \frac{\left\langle Z_{v e^{\prime} f}, Z_{v^{\prime} e^{\prime} f}\right\rangle}{\left\|Z_{v e^{\prime} f}\right\|\left\|Z_{v^{\prime} e^{\prime} f}\right\|}+i \gamma \ln \frac{\left\|Z_{v e^{\prime} f}\right\|^2}{\left\|Z_{v^{\prime} e^{\prime} f}\right\|^2} .
\ee
Here, $Z_{vef}=g_{ve}^{\rm T} z_{vf}$, and $\langle\cdot,\cdot\rangle$ denotes the $\Su$ invariant inner product. In the dual complex $\ck^*$ (denoting $v^*,e^*,f^*$ as the dual of $v,e,f$), the orientation of $\partial f^*$ is outgoing from the vertex $v^*$ and incoming to another vertex $v'^*$ (the orientation of the face $f^*$ dual to $f$ induces the orientation of $\partial f^*$). 
\\[.7em]
\emph{Spacelike triangles in timelike tetrahedra:} The spinfoam model with timelike tetrahedra is known as the Hnybida-Conrady extension \cite{Conrady:2010kc,Conrady:2010vx}. Let us restrict ourself to the case where all timelike tetrahedra are internal, that is what happens in our model. The corresponding integration variables are given by 
	\be
	\left\{g_{v e}, z_{v f}, \xi_{e h}^{ \pm}\right\}\subset X .
	\ee 
Here, $\xi_{eh}^{\pm}=v_{eh}\xi_{0}^\pm \in \mathbb{C}^2$ represents an SU(1,1) group element $v_{eh}$ that rotates $\xi_0^+ = (1,0)$ and $\xi_0^- = (0, 1)$ to $\xi_{eh}^{\pm}$. The action for spacelike triangles $h$ in timelike tetrahedra is given by \cite{Kaminski:2017eew, Han:2021rjo}:
	\be
	S_{\textit{s-t}}=\sum_{v,h\subset v} j_h  F_{vh}[X], \label{TSaction}
	\ee 
	where $F_{vf}$ is defined as follows:
\be
	F_{vh} &=& -2\ln\left[(m_{eh} \langle Z_{veh}, \xi^\pm_{eh} \rangle)(m_{e'h}\langle \xi^\pm_{e'h}, Z_{ve'h} \rangle)\right]\nonumber\\ 
	&&  + (\imath\gamma+1)\ln\left( m_{eh}\langle Z_{veh},Z_{veh}\rangle\right) \nonumber\\
    &&  + (1-\imath\gamma)\ln\left( m_{e'h}\langle Z_{ve'h},Z_{ve'h}\rangle\right).  \label{F1func} 
\ee
{The orientation of $\partial h^*$ leaves the edge $e'$ and enters the edge $e$.} Here, $\langle\cdot,\cdot\rangle$ is the $\mathrm{SU}(1,1)$ invariant inner product, and $m_{eh}=\pm 1=\langle \xi_{0}^{\pm},\xi_{0}^{\pm}\rangle$. The integration is confined to the domain where $m_{ef}\langle Z_{vef}, Z_{vef} \rangle>0$.
\\[.7em]
\emph{Timelike triangle:}  The timelike tetrahedra in our model contain timelike triangles. The corresponding integration variable are given by
	\be 
\left\{g_{v e}, z_{v f}, l_{e h}^{+}\right\}\subset X.
	\ee
Here, $l_{eh}^{\pm}=v_{eh} l_0^\pm \in \mathbb{C}^2$ represents an SU(1,1) group element that rotates $l_0^{+} = (1,1)$ and $l_0^{-} = (1,-1)$ to $l_{eh}^{\pm}$. All timelike triangle are internal in our model. Furthermore, the vertex ampliutde related to the timelike triangles gives the integrand of the form $e^{S_+}+e^{S_-}+e^{S_{x+}}+e^{S_{x-}} $ \cite{Liu:2018gfc}. Here the critical points only relate to one term $e^{S_+}$ since in our model we consider every timelike tetrahedron to have at least one spacelike triangle. $S_+$ explicitly depends on $l^+_{eh}$.
 
The action for timelike triangles is given by:
\be 
S_{\textit{t-t}} = \sum_{v,h\subset v} j_h \left(F_{ve'h}[X] - F_{veh}[X] \right)
\ee
The quantum area $\mathrm{Ar}_h$ equals the half-integer $j_h>0$. The orientation of $\partial h^*$ leaves the edge $e'$ and enters the edge $e$. $F_{veh}$ is given by
\be
F_{veh}=2\ln\sqrt{\frac{\langle Z_{veh},l^+_{eh}\rangle}{\langle l^+_{eh},Z_{veh}\rangle}}-\frac{\imath}{\gamma}\ln\langle Z_{veh},l_{eh}^+\rangle\langle l^+_{eh},Z_{veh}\rangle.
\ee
Here, $\langle\cdot, \cdot\rangle$ is the $\mathrm{SU}(1,1)$ invariant inner product on $\C^2$. All other variables remain consistent with the EPRL spinfoam. $l^-_{eh}$ does not appear in the action but appears in the critical equation as $\mathring{z}_{vf}\propto \left(\mathring{g}^{\rm T}_{ve}\right)^{-1}\mathring{l}^{-}_{ef}$.

The spinfoam action $S$ in (\ref{SFamplitude}) sums the contributions from different types of triangles:
\be
S[j_h, X; j_b, \xi_{eb}]=S_{\textit{s-s}}+S_{\textit{s-t}}+S_{\textit{t-t}}.\label{SF} 
\ee 
The continuous gauge freedom and the corresponding gauge fixings are reviewed in Appendix \ref{Gaugefix}.

The integral $\int \rmd X\, e^S$ in \eqref{SFamplitude} has been generally proven to be finite only in the case of the EPRL model with only spacelike tetrahedra. It is discussed in \cite{Han:2021bln} that the integral is possibly divergent in presence of timelike tetrahedra due to the non-compact integrals over SU(1,1) spinors $\xi_{eh}^{\pm}$. This depends on the fact that the space of SU(1,1) intertwiners is infinite-dimensional. A complete investigation of finiteness/divergence in presence of timelike triangles is still lacking in the literature. To make \eqref{SFamplitude} well-defined, we define the integral with a cut-off, ensuring that the integration domain is compact. The cut-off does not affect the analysis in this paper since we only focus on the properties of the integral within the local neighborhood at a critical point.

\subsection{Spinfoam amplitude with coherent spin-network boundary state}\label{Spinfoam amplitude with coherent spin-network boundary state}

\begin{figure}[h]       
    \includegraphics[width=0.5\textwidth]{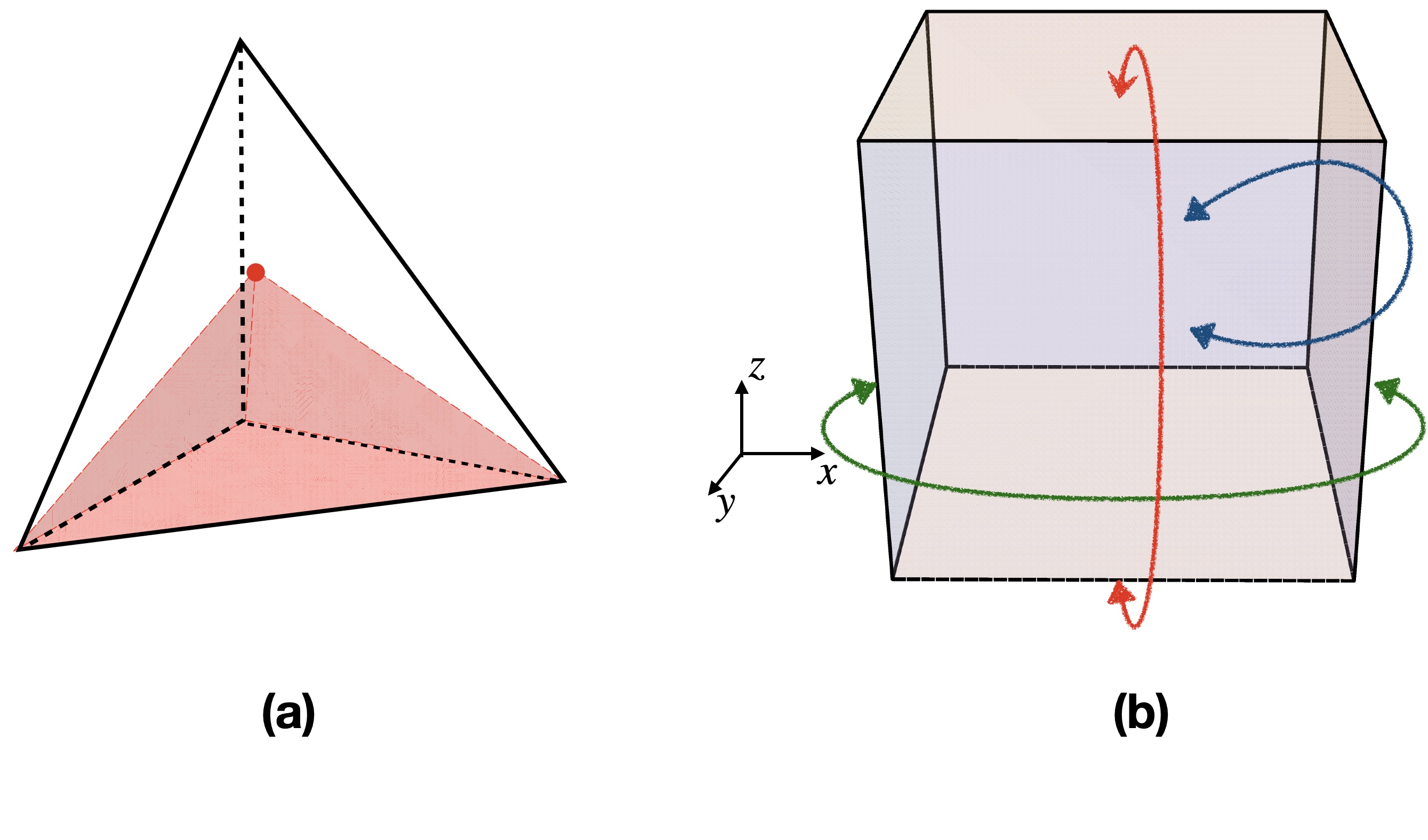}
    \caption{(a) Upon subdividing a tetrahedron $e$ into 4 tetrahedra centered around the barycentre (red point), the volume $V_b(e)$, depicted in red shading, represents the volume of the sub-tetrahedron connected to a face $b$. (b) This figure illustrates periodic boundary conditions along the spatial directions for the hypercube. The arrows indicate that faces in this direction can be identified as the same face.}
    \label{ratio}
\end{figure}

In the previous discussions, the coherent intertwiners labeled by  $j_b$ and $\xi_{eb}$ have been employed as the boundary state for the spinfoam amplitude. These characterize the boundary 3d geometry, i.e. the geometries of boundary tetrahedra. However, in this work, we are also interested in understanding the semiclassical evolution given by spinfoams with boundary data as a point in phase space. Namely, we would like to include the extrinsic curvature in the boundary data in order to compare the dynamics of the spinfoam to the 3+1 formulation of GR. For this purpose, we adopt coherent spin network states as the boundary states. Coherent spin-networks are a class of semiclassical states peaked on both intrinsic and extrinsic geometry, and can be expressed as a superposition of coherent intertwiners $|i(\vec{j}_b,\vec{\xi}_{eb})\rangle$ (see e.g., \cite{Bianchi:2009ky,Bianchi:2010mw,propagator1,propagator2,Bianchi:2011hp}, also see Appendix \ref{Coherent states in the gravitational field and the scalar field}):
\be
\left|\Psi_0\right\rangle=\sum_{j_{b}\in\Z_+/2\cup\{0\}}\psi_{j_b^0,\vartheta_b^0}(\vec{j}_b)\otimes_{a=1}^{N_b}|i_a(\vec{j}_b,\vec{\xi}_{eb}) \rangle. \label{semi-cohere}
\ee
Here, $N_b$ is the number of boundary faces, and coefficients $\psi_{j_b^0,\vartheta_b^0}(\vec{j}_b)$ are given by a Gaussian times a phase,
\be
\psi_{j^{0}_b,\vartheta^{0}_b}(\vec{j}_b) =   \exp\left[I_{j^{0}_b,\vartheta^{0}_b}(\vec{j}_b)\right]
\ee 
where
\be
I_{j^{0}_b,\vartheta^{0}_b}=-i \sum_{b} \gamma \vartheta^{0}_{b}\left(j_{b}-j^{0}_{b}\right)-\sum_{b}\frac{1}{2j^0_b} \lt({j_{b}-j^{0}_{b}}\rt)^2.\label{psiDef}
\ee 
The coherent spin-network state $\left|\Psi_0\right\rangle$ is closely related to Thiemann's complexifier coherent state in the large-$j$ regime \cite{Bianchi:2009ky,Thiemann:2000bw}. The parameters $(j_b^0,\vartheta_b^0,\xi_{eb})$ are the semiclassical boundary data of the spinfoam amplitude. Here, $j_b^0$ corresponds to the area of the boundary triangle $b$, and $\vartheta^{0}_{b}$ is the boundary dihedral angle associated with the simplicial extrinsic curvature $K$ by the relation
\be 
\vartheta^{0}_b(K) = d_b\times K_b.
\ee
In this expression, $K_b$ represents a typical value of $K$ in a neighborhood of the triangle $b$, and $K_b=3\dot{a}/a$ is a constant in the case of homogeneous and isotropic cosmology (see Appendix \ref{Kandtheta}).  The term $d_b=V_b/\mathrm{Ar}_b$ denotes the ratio of the volume $V_b$ associated with the triangle $b$ to the area $\mathrm{Ar}_b$. The triangle $b$ is shared by two boundary tetrahedra $e_1,e_2$, and $V_b=V_b(e_1)+V_b(e_2)$ is the sum of the volumes $V_b(e_{1,2})$. Each volume corresponds to a tetrahedron $e_{1,2}$, and when subdividing a tetrahedron $e$ into 4 tetrahedra according to the barycentre, $V_b(e)$ is the volume of the sub-tetrahedron connected to $b$, as illustrated in FIG. \ref{ratio}(a).

In the hypercube and double-hypercuble complex considered in this paper, two disconnected boundary components correspond to the (discrete) initial and final spatial slice, with triangles in the past and future boundaries denoted as $b_i$ and $b_f$. The spinfoam amplitude with a coherent spin-network boundary state, denoted as $A'(\mathcal{K})$, additionally sums over $j_{b}$ weighted by $\psi_{j^0_b,\vartheta^0_b}(\vec{j}_b)$ 
\begin{equation}
\begin{aligned}
A'(\mathcal{K})&:=\sum_{\{j_{b_i},j_{b_f}\}} \bar{\psi}_{j^0_{b_f},\vartheta^0_{b_f}}(\vec{j}_{b_f})\, A(\mathcal{K}) \,\psi_{j^0_{b_i},\vartheta^0_{b_i}}(\vec{j}_{b_i})  \\
&= \sum_{\{j_{b},j_h\}}  \prod_h \mu_h({j_h}) \int[\mathrm{d} X] \,e^{S_{\text{SF}}[j_f, X; j_b^0, \xi_{eb}, \vartheta_{b}]}.\label{amptot}
\end{aligned}
\end{equation}
The spinfoam action for $A'(\ck)$, denoted as $S_{\rm SF}$, is given by:
\be
S_{\rm SF}=S[j_h, X; j_b, \xi_{eb}]+I_{j^{0}_{b_i},\vartheta^{0}_{b_i}}(\vec{j}_{b_i})+\bar{I}_{j^{0}_{b_f},\vartheta^{0}_{b_f}}(\vec{j}_{b_f}),
\ee
where $S[j_h, X; j_b, \xi_{eb}]$ refers to the action defined in (\ref{SF}).

It is convenient to employ the Poisson summation formula \cite{BJBCrowley_1979} to replace the sums over $j_h$ and $j_b$ with integrals \cite{Han:2023cen, Han:2021kll, Han:2020fil}. As a result, the spinfoam amplitude $A(\mathcal{K})$ can be expressed as:
\be
\!\!\!\!\!\!A'(\ck)&=&\sum_{\{k_h\in\mathbb{Z}\}}\int\limits_{-\epsilon}^{2j^{\rm max}+1-\epsilon} 
\prod_f\mathrm{d}(2 j_{f}) \,\prod_h\mu_h(j_f)\int [\rmd X]\, e^{ S_{\rm SF}^{(k)}},\nonumber\\
&&S_{\rm SF}^{(k)}=S_{\rm SF}+4\pi i \sum_f j_f k_f,\label{integralFormAmpSF}
\ee
which holds for arbitrary $\epsilon>0$. It's important to note that we also impose the cut-off $j_b^{\rm max}$ for the boundary triangles, so that some cut-offs $j_h^{\rm max}$ resulting from the triangule inequality and $j_b$ are still maintained. The Poisson summation effectively turns all $j_f$'s into continuous variables. 

Given the area spectrum $\mathrm{Ar}_f\sim \g j_f\ell_P^2$, the regime win which the area is classical 
and $\hbar$ is small corresponds to have large spins $j_f\gg1$.  This motivates us to understand the large-$j$ regime as the semiclassical regime of $A'(\ck)$. Specifically, the large-$j$ regime of $A'(\ck)$ is defined by uniformly scaling its external parameters: the coherent state labels $j_b^0$ (for all boundary triangles $b$) and the cut-offs $j^{\rm max}$ (for all triangles) by 
\be
j^0_b\to\lambda j^0_b,\qquad j^{\rm max}\to \l j^{\rm max},\qquad \lambda\gg 1.
\ee
To study the behavior of the amplitude, we change the integration variables $j_f\to \l j_f$, resulting in $S^{(k)}_{\rm SF}\rightarrow\lambda S_{\rm SF}^{(k)}$. Then, for $\lambda\gg1$, we can analyze the integral in \eqref{integralFormAmpSF} at each $k_h$ using the stationary phase method \cite{Hormander}.

\subsection{Real and Complex Critical Points} \label{CCP}
 
By the stationary phase approximation, the dominant contributions to the integrals come from the critical points satisfying the critical equations. The critical points inside the integration domain, denoted by $\{\mathring{j}_h,\mathring{X}\}$, satsify the following critical equations from $S_{\rm SF}$:
\be
\re(S_{\rm SF})&=&\partial32{X}S_{\rm SF}=0,\label{eom1}\\
\partial_{j_f}S_{\rm SF}&=&4\pi i k_f, \qquad k_f\in\Z.\label{eom2}
\ee 
We consider the integration domain as a real manifold and refer to $\{\mathring{j}_h,\mathring{X}\}$ as the \textit{real critical point}. These real critical points relate to nondegenerate\footnote{In the model studied in this paper, every 4-simplex has both spacelike and timelike tetrahedra, so degenerate geometries are absent.} Regge geometries with a curvature constraint. The existence of a real critical point depends on the boundary conditions and may not hold for generic conditions \cite{Han:2023cen,Han:2021kll}. To address this, we apply the stationary phase approximation for complex action with parameters \cite{10.1007/BFb0074195,Hormander} and compute the \emph{complex critical points}. Here, we focus on the amplitude with $k_h=0$, and a similar analysis applies to other $k_h$.

\begin{figure}[h]
    \centering
    \includegraphics[scale=0.15]{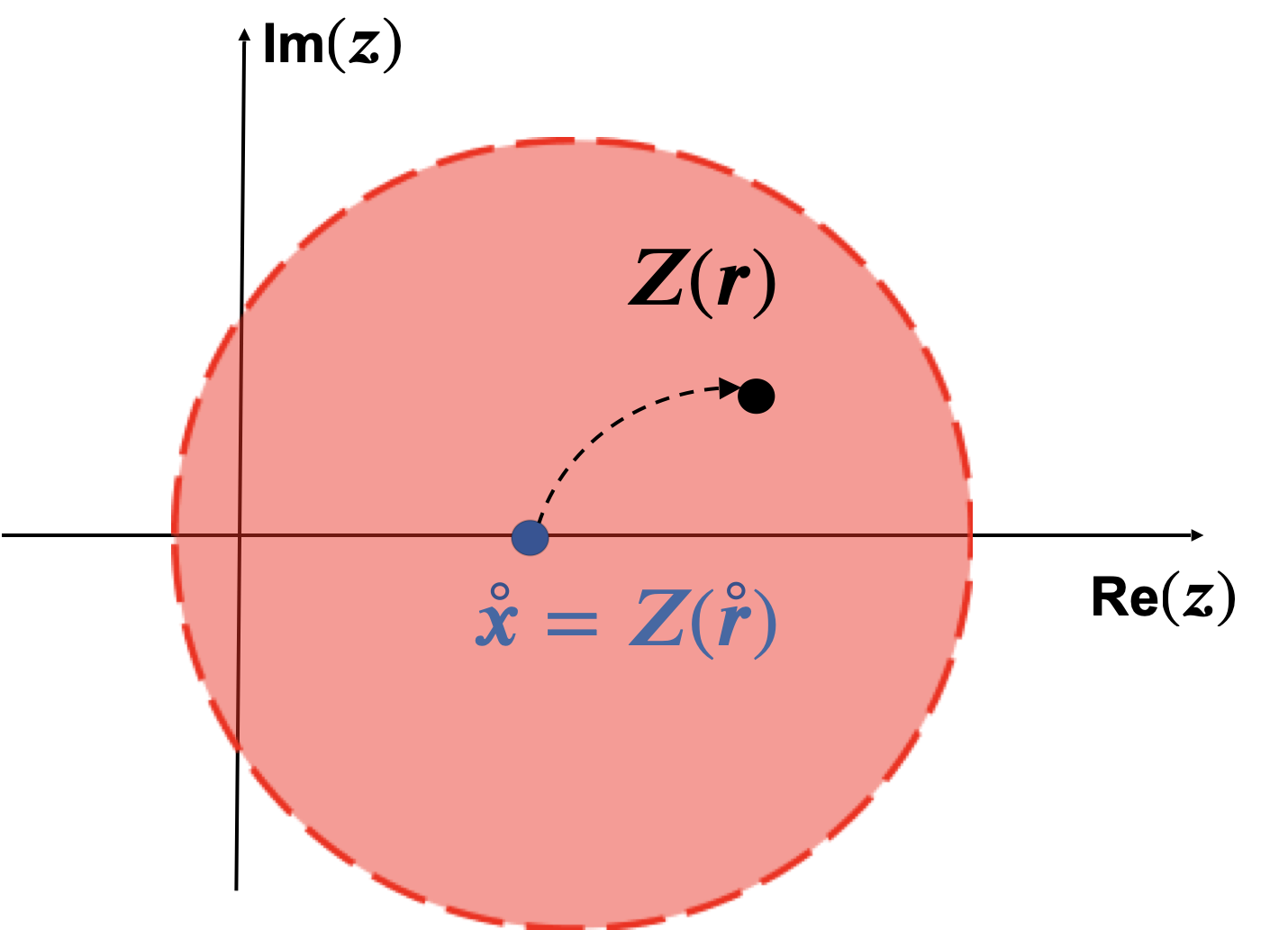}
    \caption[Caption for LOF0]{The real and complex critical points $\mathring{x}$ and $Z(r)$. $\cs(r,z)$ is analytically extended from the real axis to the complex neighborhood, as illustrated by the red disk.}
    \label{Figure0}
\end{figure}

We briefly review the analysis scheme, considering the large-$\lambda$ integral 
\be 
\int_K e^{\lambda S(r,x)}\mathrm{d}^N x,
\ee
where $r$ denotes the external parameters, and $S(r,x)$ is an analytic function of $r\in U\subset \R^k$ and $x\in K\subset \R^N$. Here, $U\times K$ forms a neighborhood of $(\mathring{r},\mathring{x})$, where $\mathring{x}$ is a real critical point of $S(\mathring{r},x)$. We denote the analytic extension of $S(r,x)$ to a complex neighborhood of $\mathring{x}$ as $\mathcal{S}(r,z)$, with $z=x+iy \in \mathbb{C}^{N}$. For generic $r\neq\mathring{r}$, the complex critical equation 
\be 
\partial_{z} \mathcal{S}=0,\label{criticaleqn111}
\ee
gives the solution $z=Z(r)$, which generically moves away from the real plane $\R^N$. Therefore, we refer to $Z(r)$ as the \emph{complex critical point} (see FIG.\ref{Figure0}). 
The large-$\lambda$ asymptotic expansion for the integral can be established with the complex critical point:
\be
&&\int_K e^{\lambda S(r,x)}  \mathrm{d}^N x \nonumber\\
&=& \left(\frac{2\pi}{\lambda}\right)^{\frac{N}{2}} \frac{e^{\lambda \mathcal{S}(r,Z(r))}}{\sqrt{\det(-\mathcal{S}_{zz}(r,Z(r)))}} \lt[1+O(1/\l)\rt].\label{asymptotics0}
\ee
Here, $\mathcal{S}(r,Z(r))$ and $\cs_{zz}(r,Z(r))= \partial^2_{z,z}\cs(r,Z(r))$ represent respectively the action and Hessian at the complex critical point. Furthermore, the real part of $\mathcal{S}$ satisfies the condition:
\be
\operatorname{Re}(\mathcal{S}) \leq-C|\operatorname{Im}(Z)|^{2}.\label{negativeReS}
\ee 
where $C$ is a positive constant. We refer to \cite{10.1007/BFb0074195,Hormander} for a detailed proof of this inequality. This condition implies that in \eqref{asymptotics0}, the oscillatory phase can only occur at the real critical point, where \,$\operatorname{Im}(Z)=0$\, and $r=\mathring{r}$. When $r$ deviates from $\mathring{r}$, causing $\operatorname{Im}(Z)$ finite and $\operatorname{Re}(\mathcal{S})$ to become negative, the result in \eqref{asymptotics0} is exponentially suppressed as $\lambda$ grows large. Nevertheless, we can reach a regime where the asymptotic behavior described in \eqref{asymptotics0} is not suppressed at the complex critical point. In fact, for any sufficiently large $\lambda$, we can always find a value of $r$ close to but not equal to $\mathring{r}$. In this region, both $\operatorname{Im}(Z)$ and $\mathrm{Re}(\mathcal{S})$ can be made small enough, ensuring that $e^{\lambda \cs}$ in \eqref{asymptotics0} is not significantly suppressed at the complex critical point. 

Recent results have clarified the significance of complex critical points in capturing curved Regge geometries 
\cite{Han:2021kll,Han:2023cen}
and 
in resolving the flatness problem (see \cite{Engle:2021xfs}). In this work, we apply these methods to the spinfoam amplitudes on a hypercube complex and double-hypercube complex, exploring their coupling with a scalar field through numerical computations. The behavior and properties of the spinfoam 
coupling to a scalar field allows us to connect with the effective dyanmics of homogeneous and isotropic quantum cosmology.

\section{Setups for Hypercube complex and double hypercube complex} \label{Complex}
This section applies the above general procedure to the spcific simplicial complexes: the hypercube complex and the double-hypercube complex  (see e.g. \cite{Dittrich:2022yoo,Han:2018fmu} for some earlier studies of spinfoams on hypercube complexes; a combinatorelly equivalent graph had been considered in spinfoam cosmology to model a 3-sphere geometry \cite{Frisoni:2023lvb}). 

\subsection{Hypercube complex and the periodic boundary condition}
\begin{figure}[h]
    \centering
	\includegraphics[scale=0.08]{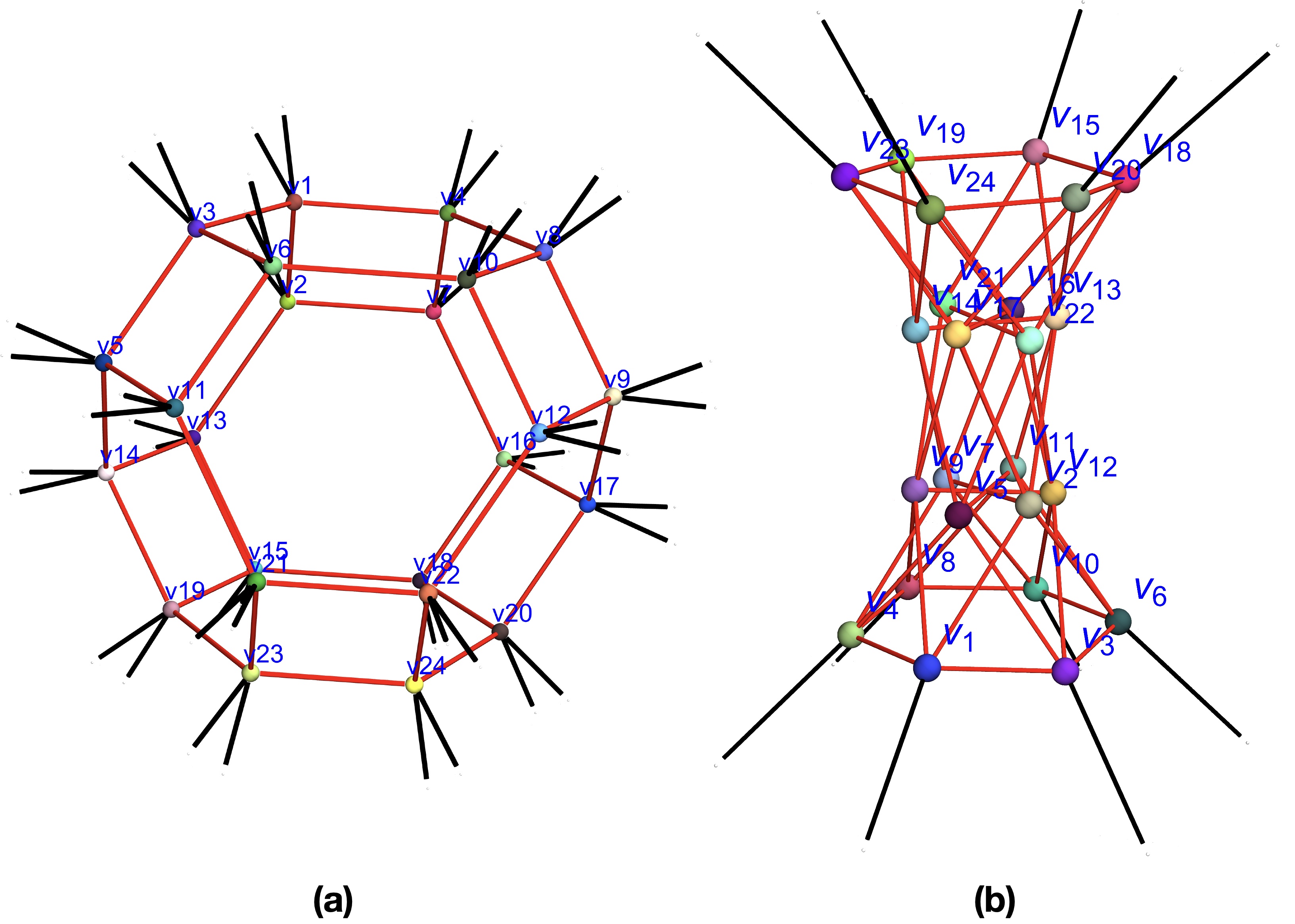} 
	\caption{(a) The dual 2-complex of the triangulation of the hypercube. (b) The dual 2-complex of the triangulation of the hypercube with spatial periodic boundary condition. In both (a) and (b), each vertex labelled by $v_i$ is dual to a 4-simplex in the triangulation, and every edge labeled by $e$ is dual to a tetrahedron. The closed loops in the diagram are dual to the internal triangles. To distinguish between boundary tetrahedra and bulk tetrahedra, we use black segments to represent boundary tetrahedra, while using red segments to represent bulk tetrahedra.}\label{hypercubedual}
\end{figure}

In 4D, a hypercube consists of sixteen vertices and eight 3D cubes on the boundary. These cubes are classifed based on their normal directions: $t, x, y, z$, and there are two cubes perpendicular to each direction. The hypercube labelled $(1,2,\cdots,16)$ has vertices with Cartesian coordinates $(t, x, y, z)$ in flat spacetime as follows:
\be
(t, x, y, z) = (0 \text{ or } h, 0 \text{ or } a, 0 \text{ or } a, 0\text{ or } a). \label{coord}
\ee 
Here, $h>0$ and $a>0$. The hypercube is triangulated into twenty-four 4-simplices. There are a total of 8 cubes forming the boundary. Each of the boundary cubes can be subdivided into six tetrahedra. We refer to Appendix \ref{1Hypercube} for details about the triangulation of the hypercube. These resulting twenty-four 4-simplices are labelled as $(v_1,v_2,\cdots,v_{24})$. The 2-complex dual to the triangulation of the hypercube is depicted in FIG. \ref{hypercubedual}(a). In this graph, each vertex labeled as $v_i$ corresponds to a 4-simplex, while each edge labeled as $e$ is dual to a tetrahedron. The closed loops in the diagram are dual to the internal triangles of the triangulation. 

\begin{figure}[h]
\centering
\includegraphics[scale=0.05]{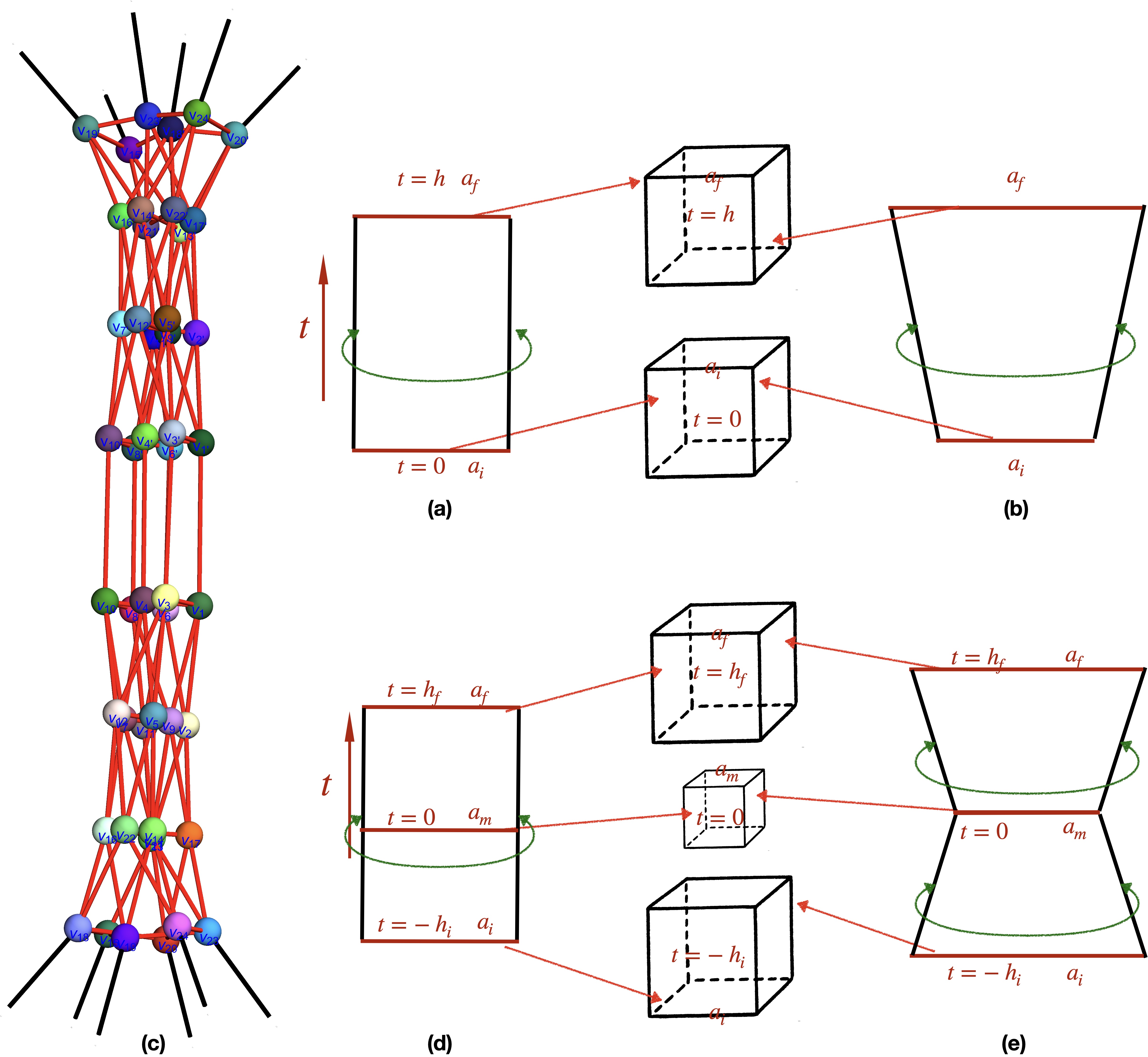}
\caption{In Figures (a) and (b), under periodic boundary conditions, the top and bottom lines represent spacelike cubes. The 4D spacetime is the region between the two boundary cubes. The cubes at $t=0$ and $t=h$ have edge lengths $a_i$ and $a_f$. Scenario (a) with $a_f=a_i$ has the flat Regge geometry, while scenario (b) with $a_f\neq a_i$ has a curved Regge geometry. Figure (c) depicts the dual diagram of the double-hypercubes. Figures (d) and (e) illustrate two hypercubes sharing a spacelike cube. The three cubes at $t=-h,0,h$ have edge lengths $a_i$, $a_m$ and $a_f$. Scenario (c) with $a_f=a_m=a_i=a$ with $\delta a=0$ represents a flat Regge geometry. The case in (d) with $a_f=a_i=a$ and $a_m<a$ exhibits a curved Regge geometry, analogous to the cosmic bounce in LQC.
}
\label{hypercube1top}
\end{figure}

We view the hypercube as a basic cell in the cosmological spacetime. The spatial boundary of the hypercube consists of a pair of spacelike cubes at $t=0,h$. Each cube is a basic cell in the spatial slice. The property that the universe is homogeneous and isotropic motivates us to impose the three-dimensional periodic boundary conditions, so that we can use a single cube to represent the entire spacial slice. The condition is imposed by matching corresponding faces along spatial directions, as illustrated in FIG. \ref{ratio}(b). The periodic boundary conditions identify 3 pairs of timelike cubes at $x=0,a$, $y=0,a$, and $z=0,a$. As a result, all timelike tetrahedra and triangles becomes internal, and all boundary tetrahedra and triangles are in two spacelike cubes at $t=0,h$, so they are all spacelike. More details about the periodic boundary condtions can be found in Appendix \ref{1Hypercube}. The resulting simplicial complex is a discretization of $I\times \mathbb{T}^3$ where $I$ is an interval and $\mathbb{T}^3$ is the 3-torus. In this work, we only focus on the flat spatial geometry corresponding to the $k=0$ cosmology. As a consequence of the periodic boundary conditions, the dual diagram representing the hypercube triangulation becomes FIG. \ref{hypercubedual}(b). In FIG. \ref{hypercube1top}(a) and (b), the top and bottom circles represent two spacelike cubes at $t=0,h$ with the periodic boundary condition. The 4D hypercube triangulation with the periodic boundary condition is composed of the following building blocks: 
\begin{itemize}
    \item 24 4-simplices.
    \item 66 tetrahedra: consisting of 54 internal tetrahedra and 12 boundary tetrahedra. Among the 54 internal tetrahedra, 36 are timelike, and 18 are spacelike. All 12 boundary tetrahedra are spacelike.
    \item 62 triangles: consisting of 38 internal triangles and 24 boundary triangles. Among the 38 internal triangles, 14 are timelike, and 24 are spacelike. All 24 boundary triangles are spacelike. 
\end{itemize}
In the flat hypercube geometry, the spacelike/timelike properties of tetrahedra and triangles are related to the ratio between $a$ and $h$. The above properties are derived from the example with $h/a=0.8$. It's worth noting that changing this ratio within a neighborhood does not affect the above properties.

The 3D Regge geometries discretizing the flat, homogeneous and isotropic spatial slices, are such that the boundary spacelike cubes at $t=h$ and $t=0$ have the lengths $a_f$ and $a_i$ respectively. Assuming both cubes are flat and equilateral fixes all edge lengths in the triangulated boundary. The 4D spacetime is the region between the boundary cubes. When the boundary edge lengths are equal $a_f=a_i$ (see FIG. \ref{hypercube1top}(a)), the hypercube is flat, and the complex is a triangulation of a 4D flat cylinder. To construct curved geometries within the hypercube complex, we need to ensure $a_f\neq a_i$ (see FIG. \ref{hypercube1top}(b)). In this context, we define $a_f = a_i - 2 \delta a$, where $\delta a\neq 0$. Before imposing the periodic boundary condition, the vertices of the future spacelike boundary cube have the coordinates
\be
t=h,\qquad x,y,z=\delta a\text{ or } a_i-\delta a,
\ee
while the vertices of the past spacelike boundary cube have the following coordinates in flat spacetime:
\be
t=0,\qquad x,y,z=0\text{ or } a_i.
\ee
The periodic boundary conditions are responsible for the presence of nontrivial curvature. For example, when $h=0.8$, $a_i=1$, and $\delta a=-0.01$, there are six non-zero deficit angles with the same value {$\delta_h=2\pi-\sum_v\theta_h(v) = 0.000625$}, where the hinges $h$ are all timelike triangles. These timelike triangles are at the place where the identification (due to the periodic boundary conditions) occurs. The deficit angles vanish at other internal triangles. $\theta_f(v)$ denotes the dihedral angle at the triangle $f$ in the 4-simplex $v$. There are nonzero boudary dehidral angles with the same value $\theta_{\rm bdry}=\sum_v\theta_b(v)=\pm 0.025$ at 12 boundary spatial triangles (6 at the future boundary and 6 at the past boundary), where the identification occurs. The dihedral angles vanish at other boundary triangles. $\theta_{\rm bdry}$ is positive for the faces in the past cube but negative for the faces in the future cube. 
Although the dihedral and deficit angles are not constant within the hypercube, the spatial homogeneity of space is realized by infintely many identical cubes, which are implemented by the above periodic boundary conditions.

\subsection{Double-hypercube complex}

To investigate a scenario analogous to the symmetric bounce in Loop Quantum Cosmology (LQC), we consider a double-hypercube. This model consists of two hypercubes: the vertices of the first hypercube are labelled by $(1,2,\cdots,8,9,\cdots,16)$, and the vertices of the second hypercube are $(1,2,\cdots,8,9',\cdots,16')$. The cube $(9, \cdots, 16)$ at $t=-h$ represents the past cube, while the future cube $(9',\cdots,16')$ is located at $t=h$. Both hypercubes share the common cube $(1,2,\cdots,8)$ at $t=0$, intuitively representing the spatial slice at the instance of cosmic bounce.

The triangulation and identification process for the second hypercube mirrors that of the first hypercube, with the only difference being the replacement of vertices $(9,\cdots, 16)$ with $(9',\cdots, 16')$ for the corresponding 4-simplices in the second hypercube. The corresponding set of 4-simplices, denoted as $(v_{1'},\cdots, v_{24'})$, follows a structure similar to (\ref{v1to24}). The dual diagram of the double-hypercube complex is shown in FIG.\ref{hypercube1top}(c). As illustrated in FIG.\ref{hypercube1top}(d), the three cubes along the time direction have the same edge length $a_f=a_m=a_i=a$, resulting in a flat geometry. To introduce curved geometries, as shown in  FIG.\ref{hypercube1top}(e), we impose $a_f=a_i=a$ but vary the edge length of the bulk cubes to 
\be 
a_m=a-2\delta a, \quad 0< \delta a<\frac{a}{2}.  \label{am}
\ee 
Here we fix $a=1$ while leaving the spatial variable $\delta a$ and the time-direction variable $h$ as parameters. The 4D double-hypercube triangulation consists of the following building blocks: 
\begin{itemize}
    \item 48 4-simplices.
    \item 126 tetrahedra, comprising 12 boundary tetrahedra and 114 bulk tetrahedra. Among the 114 bulk tetrahedra, 42 are spacelike, and 72 are timelike. All 12 boundary tetrahedra are spacelike.
    \item 112 faces, consisting of 88 bulk faces and 24 boundary faces. Among the 88 bulk faces, 60 are spacelike, and 28 are timelike. All 24 boundary faces are spacelike.
\end{itemize}
The above properties are derived from the example with $h/a=0.8$. Changing this ratio within a neighborhood does not change the spacelike and timelike properties of tetrahedra and triangles.\\

For the hypercube model shown in FIG.\ref{hypercubedual}(b), we distinguish between the boundary 4-simplices (the simplices having boundary tetrahedra) and the bulk 4-simplices (the simplices not having boundary tetrahedra). The boundary 4-simplices are:
\be
\begin{aligned}
v_{b_i}&=\left\{v_{1}, v_{3}, v_{4}, v_{6}, v_{8}, v_{10}\right\},\\
v_{b_f}&=\left\{v_{15}, v_{18}, v_{19}, v_{20}, v_{23}, v_{24}\right\},
\end{aligned}
\ee 
while other 4-simplices are considered bulk. For the double-hypercube model, the boundary 4-simplices are 
\be
\begin{aligned}
v_{b_i}&=\left\{v_{15}, v_{18}, v_{19}, v_{20}, v_{23}, v_{24}\right\},\\
v_{b_f}&=\left\{v_{15'}, v_{18'}, v_{19'}, v_{20'}, v_{23'}, v_{24'}\right\}.
\end{aligned}
\ee 

\section{Spinfoam with Scalar Matter} \label{Spinfoam with Scalar Matter}

We do not include the cosmological constant in our model.\footnote{It is possible to introduce a cosmological constant at the effective level in spinfoam cosmology \cite{Bianchi:2011ym} in strong similarity with the procedure followed here to couple the scalar field. In both cases, the contribution concerns the faces of the spinfoam, and not the edges.} \, Thus, matter coupling is needed for cosmology, as the FRW spacetime is trivially flat without matter. In the following, we discuss the spinfoam model coupled with the scalar field.

We associate a scalar $\varphi_v$ with each 4-simplex $v$. In a simplicial complex, the scalar field action is defined by: 
\be
S_{\rm L}=\frac{i}{2} \sum_{b_{v v^{\prime}}} \rho_{v v^{\prime}}\left(\varphi_v-\varphi_{v^{\prime}}\right)^2 \label{AL}
\ee  
with 
\be
\rho_{v v^{\prime}}=\sgn(b_{v v^{\prime}}) \frac{\left|V_{v v^{\prime}}\right|}{\left|b_{v v^{\prime}}\right|} \label{rho},
\ee 
where $v$ and $v'$ are neighboring 4-simplices sharing a tetrahedron. The definition of $S_{\rm L}$ is inspired by \cite{REN1988661}. Here, $b_{vv'}$ denotes the edge dual to the tetrahedron shared by $v$ and $v'$. $\text{sgn}(b_{v v^{\prime}}) = -1$ for a timelike $b_{vv'}$ and $\text{sgn}(b_{v v^{\prime}}) = 1$ for a spacelike $b_{vv'}$. Also,  $|V_{vv^{\prime}}|>0$ is (the absolute value of) the volume of the tetrahedron dual to $b_{v v^{\prime}}$, and $|b_{v v^{\prime}}|>0$ is (the absolute value of) the length associated with  $b_{vv'}$\footnote{Here, the absolute value $|x|$ is actually $\sqrt{|x^2|}$, where $x^2$ is the squared volume or squared length and can be positive or negative for being spacelike or timelike.}: 
\be 
|b_{vv^{\prime}}| = |b_{ve}| + |b_{v^\prime e}|. 
\ee 
where $|b_{ve}|>0$ is the distance from the centroid of the 4-simplex $v$ to the centroid of the tetrahedron $e\subset \partial v$. The centroid of a finite set of $k$ points $\mathbf{x}_1, \mathbf{x}_2, \ldots, \mathbf{x}_k$ in $\mathbb{R}^N$ is given by:
\be
\mathbf{C}=\sum_{i=1}^k\frac{\mathbf{x}_i}{k}.
\ee

Our main purpose is to study homogenous and isotropic cosmology. The basic cell of the cosmological spacetime is the hypercube, whose geometry is determined by the data $(a_i,a_f,h)$. Therefore, for our purpose, it is sufficient to couple the scalar field only to the geometrical data $(a_i,a_f,h)$ rather than all edge lengths in the triangulation. By using the coordinates of the 4-simplex vertices, we express $\rho_{v, v'}$ in terms of $(a_f, a_i, h)$ in the hypercube complex and $(a_f, a_i, h_f, h_i, \delta a)$ in the double-hypercube complex. See Appendix \ref{Lg} for some examples.

The coupling the scalar field to the spinfoam is defined by adding $S_{\rm L}$ to $S_{\rm SF}$ and integrating/summing both spinfoam and scalar degrees of freedom. Since the scalar couples to the lengths $a_i,a_f,h$, but the spinfoam depends on the areas $\g j_f$, we change variables from a subset of areas to recover some lengths, including $a_i,a_f,h$, by inverting Heron's formula. This approach is similar to what has been proposed earlier in \cite{Han:2021kll,Han:2023cen}. Inverting Heron's formula can be done locally in the space of areas, and the local inversion is sufficient for our purpose, as our computation is based on the continuous deviation of $\delta a$ from zero.

The Hessian of the spinfoam action turns out to be degenerate (at the real critical point for the flat geometry). So we separate some variables in the integral \eqref{integralFormAmpSF} such that the integration of the remaining variables can be studied using the stationary phase method without involving a degenerate Hessian. If we denoted by $\chi$ the set of variables for the degenerate directions of the Hessian and by $\tilde{\chi}$ the rest of the variables, the integral in \eqref{integralFormAmpSF} is written in the form $\int \rmd \chi\rmd \tilde{\chi}\, f(\chi,\tilde{\chi})$. We will only study the integral $\int\rmd \tilde{\chi}\, f(\chi,\tilde{\chi})$ using  the stationary phase method and only consider $\chi$ as parameters. 

$h$ or $h_{f(i)}$ relates to a degenerate direction \footnote{Two flat hypercubes (with periodic boundary condition) can have lengths $(a,h)$ and $(a,h+\delta h) $ with arbitrarily small $\delta h$. It implies there are continuously many real critical points of the model corresponding to flat hypercubes with different $h$, resulting in the degeneracy of the Hessian.}. Therefore, after the above change of variables, $h,h_{f(i)}\in \chi$. We have to separate $h$ or $h_{f(i)}$ from the integral and view them only as parameters. For this technical reason, $h$ or $h_{f(i)}$ is not an ideal variable for the scalar action. Here, we make a change of variable to use the hyper-diagonal $L$ or $L_{f(i)}$ instead ($L_i,L_f$ are hyper-diagonal of the hypercubes in the past and future). In the hypercube complex, we replace $h$ by 
\be
h = \sqrt{3 (a_i-\delta a)^2-L^2},
\ee
in the scalar action. In the double-hypercube, we have the replacement
\be
h_{i(f)} = \sqrt{3 (a_{i(f)}-\delta a)^2 - L_{i(f)}^2}
\ee
As a result, the action for the scalar field in (\ref{AL}) is a function of geometry variables $\boldsymbol{g} =(a_f, a_i, L)$ or $ (a_{f(i)}, \delta a, L_{f(i)})$.

Furthermore, we consider the coherent state $\psi_{z}$ as the boundary state of the scalar field \cite{Han:2021cwb}. $\psi_{z}$ is expressed as a function of $\varphi_{v_b}$, the scalar associated with the 4-simplex connecting to a boundary tetrahedron, denoted by $v_b$ \footnote{We have the overlap $\langle \phi_{z'}|\phi_z\rangle=\exp[\frac{1}{2\hbar}(\bar{z}'\cdot z-\frac{1}{2}\bar{z}'\cdot z'-\frac{1}{2}\bar{z}\cdot z)]$ and the over-completeness $\int\prod_{v}\frac{\rmd^2 z_v}{\pi\hbar}\psi_z(\varphi)\overline{\psi_z}(\varphi')=\delta (\varphi-\varphi')$, where $z=(z_v)_v$ and $\rmd^2 z=\rmd\mathrm{Re}(z)\rmd \mathrm{Im}(z)$.}:
\be 
\psi_{z} =\prod_{v_b} \exp\left(\frac{z_{v_b}^2-2\left(\varphi_{v_b}-z_{v_b}\right)^2-z_{v_b}\bar{z}_{v_b}}{4\hbar}\right).
\ee 
Here $z_{v_b}$ and $\bar{z}_{v_b}$ provide the complex parametrization of the scalar sector in the phase space:
\be 
z_{v_b}=\phi_{v_b}+i \pi_{v_b}, \quad \bar{z}_{v_b}=\phi_{v_b}-i \pi_{v_b},
\ee
with the canonical conjugate variables of the real scalar field $\phi_{v_b}$ and $\pi_{v_b}$. Therefore, the scalar field action with the coherent state as the boundary state can be expressed as 
\be 
&&S_{\rm Scalar}(\boldsymbol{g},\varphi_v;\phi_{v_{b_{i(f)}}},\pi_{v_{b_{i(f)}}}) = \frac{i}{2} \sum_{b_{v v^{\prime}}} \rho_{v v^{\prime}}\left(\varphi_v-\varphi_{v^{\prime}}\right)^2\nonumber\\
&&\qquad+\frac{1}{4\hbar}\sum_{v_{b_i}}\left(z^2_{v_{b_i}}- 2\left(\varphi_{v_{b_i}}-z_{v_{b_i}}\right)^2- z_{v_{b_i}}\bar{z}_{v_{b_i}} \right)\label{Sscalar}\\
&&\qquad + \frac{1}{4\hbar}\sum_{v_{b_f}}\left(\bar{z}^2_{v_{b_f}}-2\left(\varphi_{v_{b_f}}-\bar{z}_{v_{b_f}}\right)^2-z_{v_{b_f}}\bar{z}_{v_{b_f}}\right)\nonumber,
\ee
where the initial and final scalar data are
\be 
z_{v_{b_i}}=\phi_{v_{b_i}}+i \pi_{v_{b_i}}, \quad 
z_{v_{b_f}}=\phi_{v_{b_f}}+i \pi_{v_{b_f}}.
\ee
Here, $\phi_{v_{b_{i(f)}}}, \pi_{v_{b_{i(f)}}}$ are the boundary data that should be assigned in the numerical computation.

\section{Implementation and Numerical Results} \label{Implementation and Numerical Results}


In this section, we numerically investigate the spinfoam amplitude coupled with scalar matter on the single hypercube and on the double-hypercube complexes. The spinfoam amplitude coupled with scalar matter can be expressed as in \cite{Han:2021kll,Han:2023cen}:
\be 
\int \prod^{N_{\rm out}}_{I=1} \mathrm{~d} j^{\text{out}}_I \mathcal{Z}_{\mathcal{K}}\left(j^{\text{out}}_I,\xi_{eb},K_{i(f)},\phi_{i(f)},\pi_{i(f)}\right). \label{Z1}
\ee 
For the spinfoam sector, we only focus on the integral with $\vec{k}=0$ in \eqref{integralFormAmpSF}. We have separate some integration variables $j_I^{\rm out}$ due to the degeneratcy of the Hessian at the real critical point (of the flat geometry). The integral of $\mathcal{Z}_{\mathcal{K}}$ can be studied by the stationary phase method with a nondegenerate Hessian. $\mathcal{Z}_{\mathcal{K}}$ has the following form:
\be 
\mathcal{Z}_{\mathcal{K}}= \int \mathrm{~d}^{N}\mathbf{x} \,\mu(\mathbf{x})\, e^{S_{\rm tot}(r, \mathbf{x})}, 
\ee 
where the integration variables $\mathbf{x}$ include $j_f=(j_h,j_b)$ other than $j^{\rm out}_I$, other spinfoam variables $\xi_{eh}^\pm,l^+_{eh},z_{vf},g_{ve}$, and the scalar $\varphi_S$. The total action $S_{\rm tot}$ sums the spinfoam action and the scalar action (see Appendix \ref{ReggeWithAL} for an analog in Regge calculus):
\be 
S_{\rm tot}(r, \mathbf{x}) = S_{\rm SF} + S_{\rm Scalar}. \label{Stot}
\ee 
The parameters $r$ include the variables that are not integrated in $\cz_{\ck}$: 
\be
r=\{j^{\text{out}}_I,j^0_b,\xi_{eb},K_{b},\phi_{v_b},\pi_{v_b}\}.
\ee 

Recall the large-$j$ scaling described at the end of Section \ref{Spinfoam amplitude with coherent spin-network boundary state}. When we scale $j_f\to\l j_f$, the edge lengths $a_{i},a_f,a_m,h,L_i,L_f$ are scaled by $\sqrt{\l}$, then $\rho_{vv'}$ in $S_{\rm Scalar}$ is scaled by $\l$. Moreover, the scaling $j_f\to\l j_f$ ($\l\gg1$) comes from the area spectrum $\mathrm{Ar}_f\sim \g j_f\ell_P^2$ and the scaling $\hbar\to \l ^{-1}\hbar $. Taking the scaling of $\hbar$ into account, the total action is homogeneous under the scaling
\be
S_{\rm tot}\to \l S_{\rm tot}.
\ee
By replacing $S_{\rm tot}$ by $\l S_{\rm tot}$, the integral of $\cz_\ck$ becomes the standard form suitable for the stationary phase analysis. 

In the following, we investigate numerically $\mathcal{Z}_{\mathcal{K}}$ using the stationary phase method outlined in Section \ref{CCP}. The integration of $j_I^{\rm out}$ for the degenerate directions is not studied in this work and will be part of future investigations. The degenerate directions relate to the continuous variations of the real critical points, which all correspond to flat Regge geometries. Therefore, the degeneracy of the Hessian is expected to relate to the semiclassical diffeomorphisms, which map between flat Regge geometries.


\subsection{Single-hypercube model}

In the hypercube model, the Hessian matrix of $S_{\rm tot}$ has 4 degenerate directions at the real critical point when taking into account all integration variables. Therefore we begin by setting $N_{\rm out}=4$ in (\ref{Z1}), where the four external spins are defined as:
\be 
j_I^{\text{out}}=\{j_{1,9,14},j_{1,11,16}, j_{1,2,16},j_{1,10,16}\}.\label{joutcube}
\ee
Here, the numbers represent  vertices of the simplicial complex. Fixing these spins ensures the nondegeneracy of the Hessian with respect to $\mathbf{x}$ at both real and complex critical points. Note that the fixed spins include the length of the hypercube along the time-direction, $h$, which is a degenerate direction.
\be 
j_{1,9,14}&=&a_f\sqrt{\frac{4h^2-(a_f-a_i)^2}{8}},\nonumber\\
\gamma j_{1,11,16}= \gamma j_{1,10,16} &=& a_f \sqrt{\frac{(a_f+a_i)^2 -4h^2}{8}},\nonumber\\
\gamma j_{1,2,16} &=& a_i \sqrt{ \frac{(a_f+a_i)^2-2h^2}{8}}, \label{jandh}
\ee 
where $j_{1,9,14}$ is associated with a timelike triangle.

The scalar action $S_{\rm Scalar}$ depends on $(a_i, a_f, L, \varphi_{v})$ with $v=1,2,\ldots, 24$. By identifying $\gamma j_f$ with the area of triangle $f$ (in Planck units), we use the Heron’s formula to express  $(j_{1,2,8},j_{9,4,16},j_{1,4,16})$ in terms of the length variables $(\mathfrak{a}_i,\mathfrak{a}_f,\mathfrak{l})$ in the scalar field as follows:
\be 
j_{1,2,8}& =& \frac{\mathfrak{a}_i^2}{\sqrt{2}}, \quad j_{9,14,16} = \frac{\mathfrak{a}_f^2}{\sqrt{2}},\\
j_{1,4,16}& =& \mathfrak{a}_i\sqrt{\frac{\mathfrak{l}^2}{2}- \lt(\frac{\fa_i+\fa_f}{2}\rt)^2}.\label{j128and}
\ee 
Here, $\mathfrak{a}_i,\mathfrak{a}_f,\mathfrak{l}>0$ can be uniquely determined by $j$'s. Geometrically,  $({\sqrt{\gamma}}\frak{a}_i,{\sqrt{\gamma}}\frak{a}_f,{\sqrt{\gamma}}\frak{l})$ corresponds to the lengths $({a_i}, {a_f}, {L})$. However, we emphasize that $\fa_i,\fa_f$ are the integration variables derived from the above $j$'s, and they are distinguished from the external parameters $a_i,a_f$ that enter the coherent state data $j_b^0$ and external spins $j_I^{\rm out}$.

The integrals in $\mathcal{Z}_{\mathcal{K}}$ are $N=1192$ dimensional,
\be 
\begin{aligned}
\mathcal{Z}_{\mathcal{K}} = \int \mathrm{~d}^{N}\mathbf{x} \,|\det\mathbf{J}_{\mathbf{f}}(\frak{a}_i,\frak{a}_f,\frak{l})|\,\mu(\mathbf{x})\, e^{\lambda S_{\rm tot}(r, \mathbf{x})}, \label{Z2}
\end{aligned}
\ee
where $\mathbf{J}_{\mathbf{f}}(\frak{a}_i,\frak{a}_f,\frak{l})$ is the Jacobian for changing variables from $(j_{1,2,8}, j_{9,4,16}, j_{1,4,16})$ to $(\frak{a}_i,\frak{a}_f,\frak{l})$. The integrated variables are 
\be 
\mathbf{x}=\{g_{ve}, z_{vf},  \xi^{\pm}_{eh},l_{eh}^+, j_{\bar{h}}, \frak{a}_i,\frak{a}_f,\frak{l}, \varphi_v\}. 
\ee
Here, $j_{\bar h}$ represents  spins other than the ones in \eqref{joutcube} and \eqref{j128and}. The integration variables in $\mathbf{x}$ are subject to some gauge fixings and have the parametrizations discussed in Appendix \ref{Gaugefix}. In (\ref{Z2}), our focus is on the contribution from the (complex) neighborhood enclosing a single real critical point corresponding to the flat hypercube (the one shown in FIG.\ref{hypercube1top}(a)).

\begin{figure}[h]
    \centering
    \includegraphics[scale=0.08]{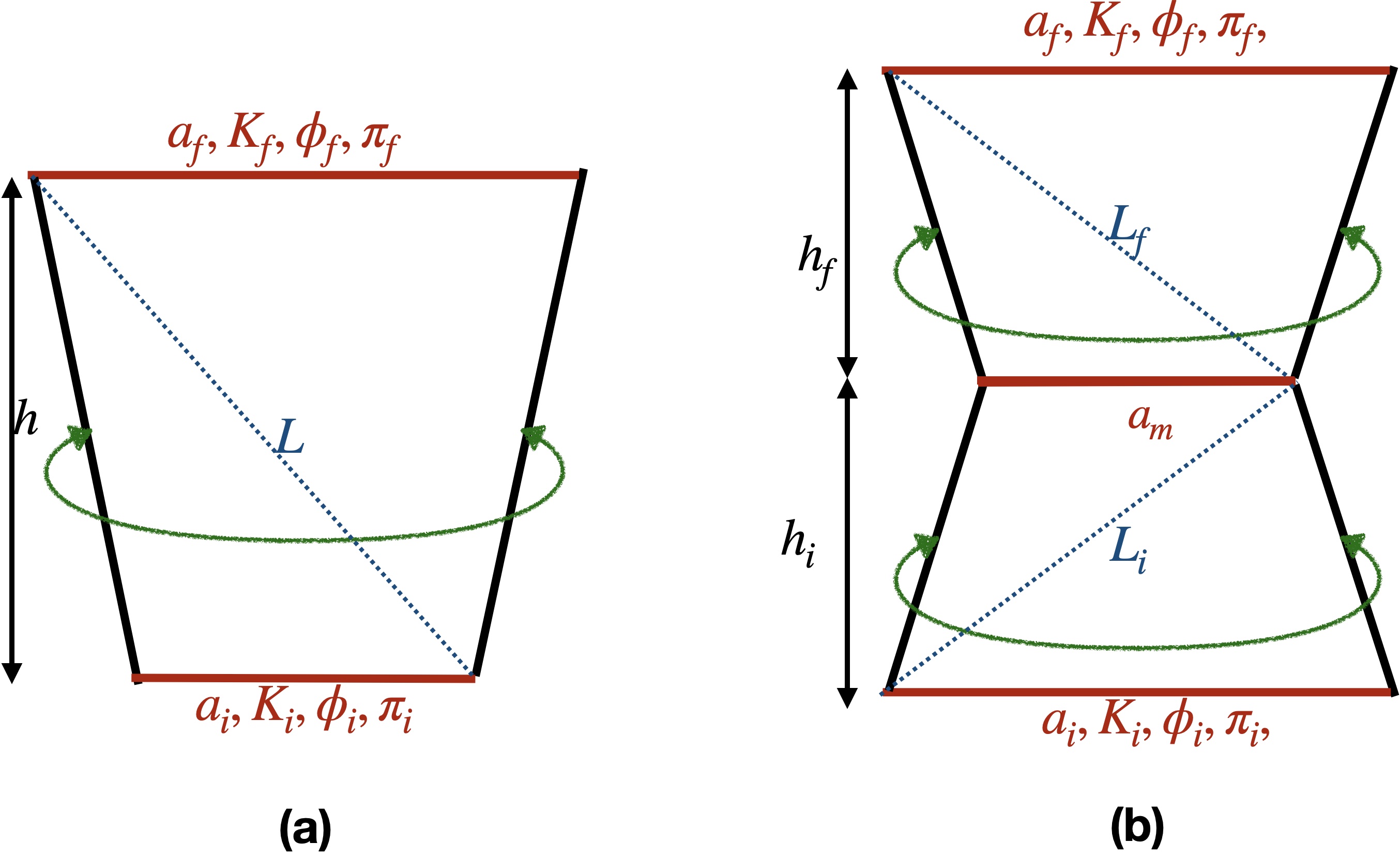}
    \caption{(a) External parameters in the hypercube model and (b) External parameters in the double-hypercube model.}
    \label{Oneand2Hypercube}
\end{figure} 

The real critical point for the flat geometry, denoted by $\mathring{\mathbf{x}}$, relates to the external data $\mathring{r}$ determined by $a_i=a_f=1$, a fixed value of $h\in[0.4,0.8]$, and $K_b=\phi_{v_b}=\pi_{v_b}=0$. It is important to note that $a_i$, $a_f$, and $K_b$ detemine the coherent state data $j_b^0,\vartheta_b^0$, while $a_i$, $a_f$ and $h$ determine $j_I^{\rm out}$. Deforming $r$ away from $\mathring{r}$ will lead the critical point to become complex. The external parameters $r$ that determine the complex critical point are described as follows: 


We set the inital data on the $t=0$ slice to satisfy the Friedmann equation:
\be
\left(\frac{\dot{a}_i}{a_i}\right)^2=\frac{8\pi G \rho_i}{3}, \quad \text{or}\quad 
K_i^2 = \frac{3\pi_i^2}{2a_i^6},
\ee 
where we set $8\pi G=1$ and denote $K_i\equiv K_{b_i}$, $\pi_i\equiv \pi_{v_{b_i}}$. We have used the scalar density $\rho_i=\frac{\pi_i^2}{2V_i^2}=\frac{\pi_i^2}{2a_i^6}$ and the extrinsic curvature $K_i=\frac{3\dot{a}_i}{a_i}$.  In our numerical computation, we set: 
\be 
a_i=1,\quad \phi_i=1,\quad \pi_{i}=\frac{1}{100}, \quad K_i = \sqrt{\frac{3}{2}}\times\frac{1}{100}. \label{InitialData}
\ee 

On the final slice $t=h$, the final data $(a_f,K_f,\phi_f, \pi_f)$ are given with the condition $a_f>a_i$, $\phi_f>\phi_i$, and $K_f,\phi_i>0$. We consider $a_f>a_i$ for the expanding universe (see FIG.\ref{Oneand2Hypercube}(a)). $\phi_f>\phi_i$ is due to $0<\pi_i\propto \dot{\phi}$. When fixing the initial data, the amplitude $\cz_\ck$ becomes a function on the phase space $\calp_f$ of final data $(a_f,K_f,\phi_f, \pi_f)$. We are interested in the local maximum of $|\cz_\ck|$ on $\calp_f$ for large $j$, expecting the final data at the maximum to be related to the Friedmann equation (the Hamiltonian constraint). In addition to the final data, other external paramters in $r$ of $\cz_\ck$ include $h$. We can compute the large-$j$ approximation of $\cz_\ck$ with fixed $r$, following the scheme in Section \ref{CCP}. However, it is technically difficult to scan the entire space of $r$ and search for the maxima, given that the space of $r$ is high-dimensional. In our computation scheme, we fix $a_f,\phi_f,h$ ($j_I^{\rm out}$ are determined by $a_f,a_i,h$) and restrict $\cz_\ck$ to a function of two variables $K_f,\pi_f$. Then, we compute $\cz_\ck$ for a large number of $(K_f,\pi_f)$ samples and find the maxima of $|\cz_\ck|$. There is only a subspace of  $(K_f,\pi_f)$ where $|\cz_\ck|$ reaches the maxima. The subspace turns out to satisfy a constraint equation, which is understood as a modified Friedmann equation emerging from the spinfoam coupled to scalar matter. In pactice, we first set $a_f=1.02$, $\phi_f=1.04$, $h=0.8$, and numerically compute the constraint equation for the $(K_f,\pi_f)$. We then vary the values of $\phi_f$ and $h$ to investigate how they affect the constraint equation. We fix $\gamma=10^{-2}$ throughout the entire computation.

In (\ref{Z2}), both $S_{\rm tot}(\mathbf{x})$ and $\mu(\mathbf{x})$ are analytic in the neighborhood $K$ of the real critical point $\mathring{\mathbf{x}}$. We extend $S_{\text{tot}}(r, \mathbf{x})$ to a holomorphic function $\mathcal{S}_{\text{tot}}(r,\mathbf{z})$ ($\mathbf{z}\in \mathbb{C}^{1192}$) in a complex neighborhood\footnote{A formal discussion of the analytic continuation of the spinfoam action is provided in \cite{Han:2021rjo}.}. Given the values of $a_f,a_i,h$ and the sample $(K_f,\pi_f)$, we find the numerical solutions to the complex critical equations
\be 
\partial_{\mathbf{z}} \mathcal{S}_{\text{tot}}(r, \mathbf{z})=0. \label{CCPeqns}
\ee
The solution $Z(r)$ is complex. We evaluate the total action $\mathcal{S}_{\text{tot}}(r,Z(r))$ at the complex critical point. The large-$j$ approximation of $\cz_{\ck}$ is given by \eqref{asymptotics0}. 

\begin{figure}[h]
    \centering
    \includegraphics[scale=0.06]{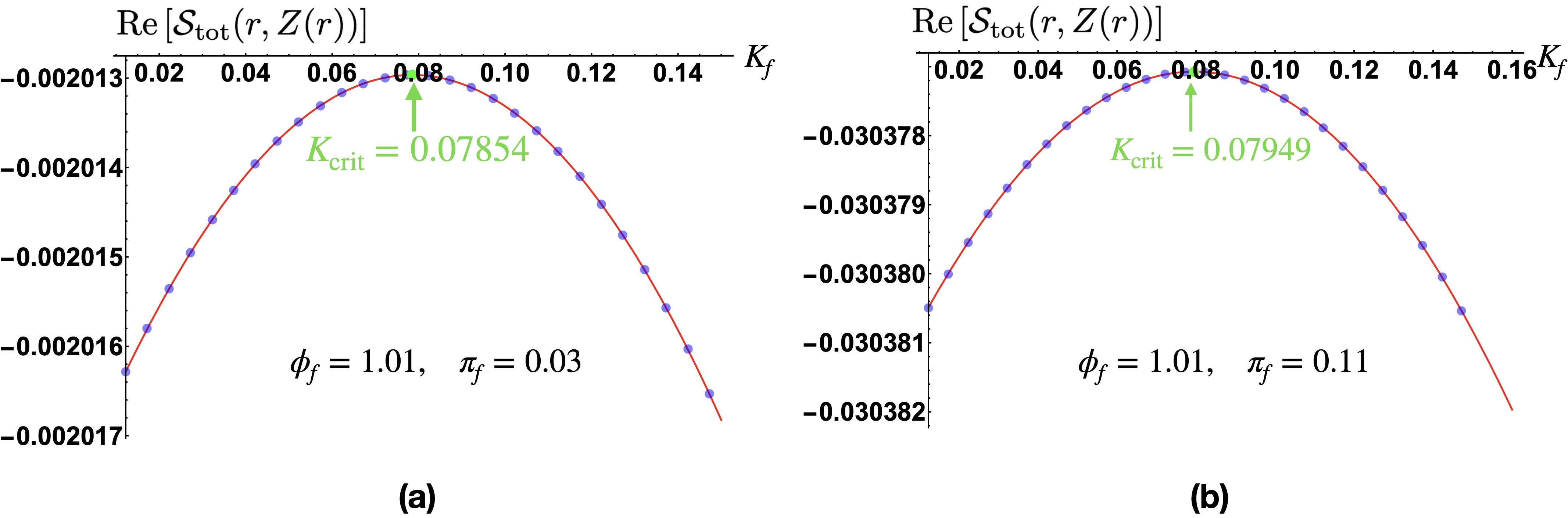}
    \caption{The real part of the total action $\mathcal{S}_{\text{tot}}(r,Z(r))$ at the complex critical point $Z(r)$ as a function of the extrinsic curvature $K_f$ on the final slice $t=h=0.8$. Two different values of $\pi_f$ are considered in (a) $\phi_f=1.01, \pi_f=0.03$, and in (b) $\phi_f=1.01, \pi_f=0.11$. The red curves are approximated functions that interpolate the data points (blue) on the plot. The maximum is located at $K_{\rm crit}$ (green),  determined by the interpolation function.}
    \label{ReSvsK}
\end{figure}

Fixing a value of $\pi_f$, we perform the numerical computation $\mathcal{S}_{\text{tot}}(r,Z(r))$ for samples of $K_f$ values, covering small to large extrinsic curvature. FIG.\ref{ReSvsK} plots the results of $\Re\left[\mathcal{S}_{\text{tot}}(r,Z(r)) \right]$ at $\pi_f=0.03$ (in FIG.\ref{ReSvsK}(a)) and $\pi_f=0.11$ (in FIG.\ref{ReSvsK}(b)). By the numerical interporlation, the red curves provide smooth fits to the data points, enabling us to identify the maximum at $K_f=K_{\rm crit}$. The maxima of $\Re[\mathcal{S}_{\text{tot}}(r,Z(r))]$ and $|\cz_\ck|$ are approximately located at the same point\footnote{We may write the right-hand side of \eqref{asymptotics0} as 
\[
\exp[\l \cs -\frac{N}{2}\ln(\frac{\l}{2\pi})-\ln(\sqrt{\det(-\cs_{zz})})+O(1/\l)],
\] 
where $\lambda \cs$ dominates the exponent for large $\l$. So the critical point of the exponent's real part is approximated by the critical point of $\mathrm{Re}(\cs)$.} (green color in FIG. \ref{ReSvsK}). The location of $K_{\rm crit}$ varies depending on the values of $\pi_f$. Nevertheless, given any $\pi_f>0$, the algorithm outlined above consistently provides the corresponding $K_{\rm crit}$.

\begin{figure}[h]
    \centering
    \bigskip
    \includegraphics[scale=0.12]{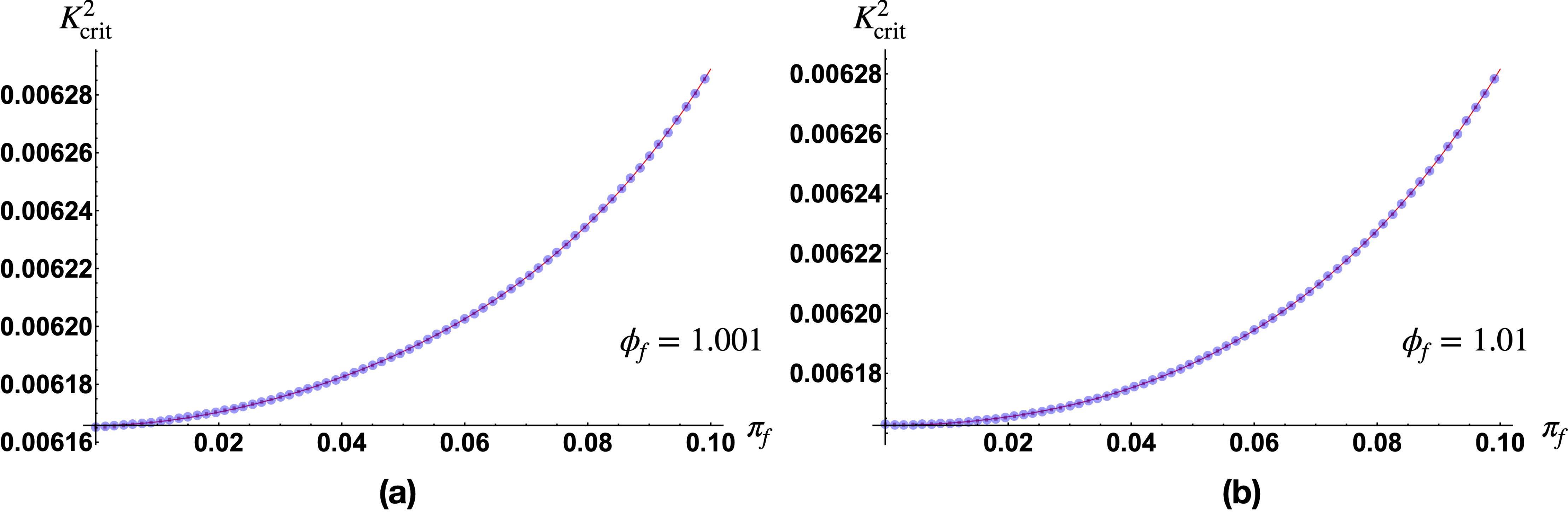}
    \caption{The variation of $K^2_{\rm crit}$ with respect to $\pi_f$ is shown in (a) for  $\phi_f=1.001$ and in (b) for $\phi_f=1.01$. The error bars, obtained from the default error estimation using "\textbf{NonlinearModelFit}" in \textit{Mathematica 13.0.1.0}, are represented by the red fences, though they may be too small to be clearly visible. }
    \label{KvsPi}
\end{figure}

\begin{figure}[h]
    \centering
    \includegraphics[scale=0.115]{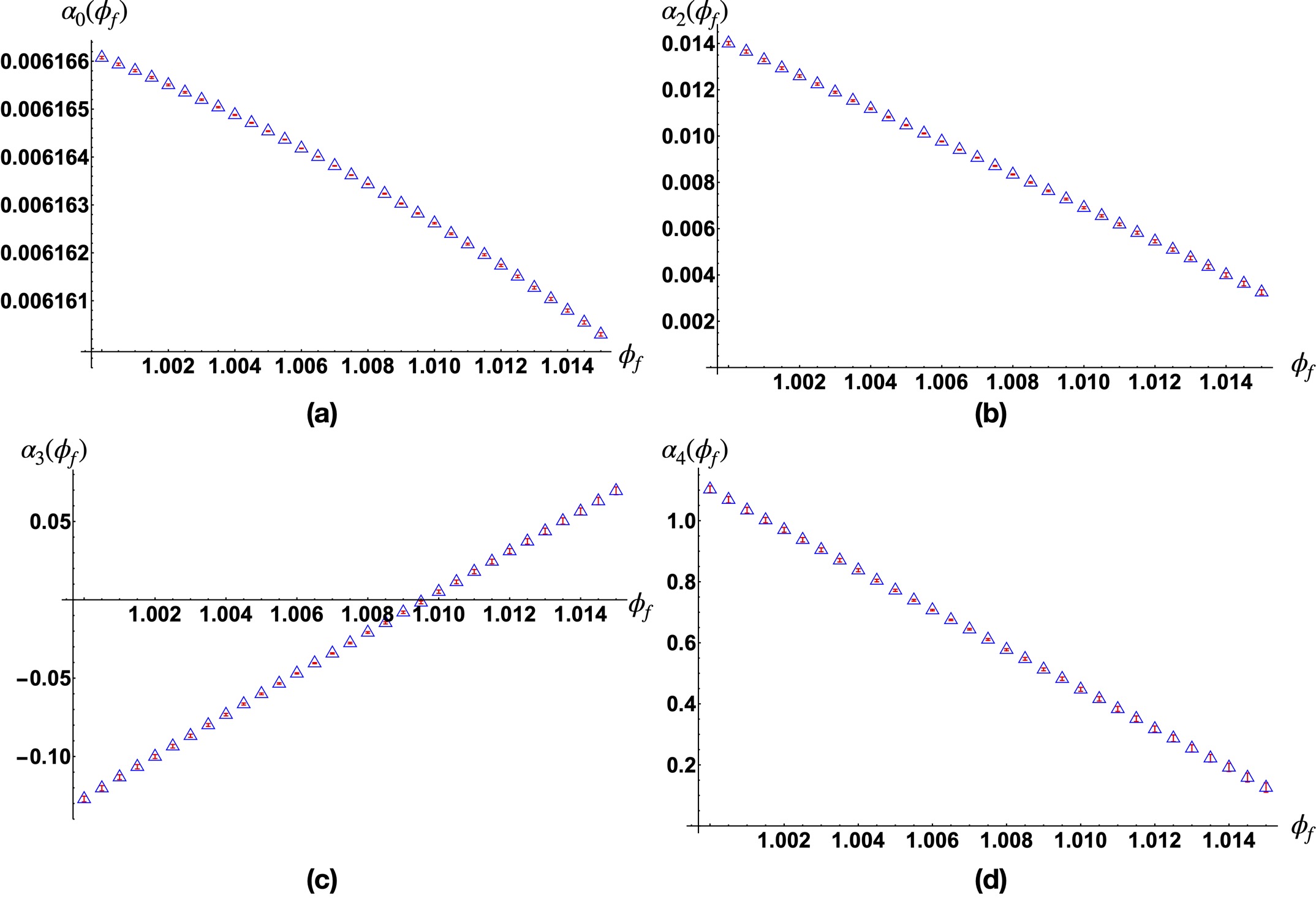}
    \caption{The relationship between $K_{\rm crit}^2$ and $\phi_f$ using the model (\ref{a012}) at $h=0.8$. The blue uptriangles represent the variations of $\alpha_0$,$\alpha_2$, $\alpha_3$, $\alpha_4$ with $\phi_f$. The error bars are denoted by the red fences.}
    \label{ParamsvsPhi2}
\end{figure}

By varying $\pi_f$, the numerical computations yield data points (depicted as blue points in FIG.\ref{KvsPi}) demonstrate the relationship between $K^2_{\rm crit}$ and $\pi_f$. The constraint equation of $K_f$ and $\pi_f$ is given by finding a smooth function to fit the data points. We assume that the relationship between $K^2_{\rm crit}$ and $\pi_f$ can be effectively modeled by a polynomial expression of the form:
\be 
K^2_{\rm crit} = \alpha_0 + \alpha_2\pi^2_f+\alpha_3\pi^3_f+\alpha_4\pi^4_f +O(\pi^5_f).\label{a012}
\ee
We employ ``\textbf{NonlinearModelFit}'' in \textit{Mathematica 13.0.1.0} and use the default unbiased estimate of errors to determine the coefficients $\alpha_0, \alpha_2, \alpha_3,\alpha_4$ from the numerical data. The linear term in $\pi_f$ is excluded from \eqref{a012} based on a comparision of errors between the nonlinear fit with \eqref{a012} and one including the linear term. We observe that the model with the linear term produces significantly larger errors than FIG. \ref{ParamsvsPhi2}. Therefore, we adopt (\ref{a012}) to represent the relationship between $K_{\rm crit}^2$ and $\pi_f$. 


Note that since the above algorithm relies on fixed values of $a_f$, $\phi_f$ and $h$, the coefficients $\alpha_0, \alpha_2, \alpha_3,\alpha_4$ depend on the choices of $a_f,\phi_f,h$, i.e.,
\be
\alpha_i=\alpha_i(a_f,\phi_f,h),\qquad i=0,2,3,4.
\ee

Specifically, for the case in FIG.\ref{KvsPi}(a) where $\phi_f=1.001$ and $a_f,h$ are the same as above, the parameters with errors are:
\be 
\begin{aligned}
\alpha_0 &= 0.00617_{\pm 1.08\times 10^{-8}},\quad \alpha_2 = 0.0133_{\pm 6.32\times 10^{-5}},\\
\alpha_3 &= -0.113_{\pm 1.52\times 10^{-3}},\quad \alpha_4 = 1.034_{\pm 9.39\times 10^{-3}}.
\end{aligned}
\ee  
For $\phi_f=1.01$ in FIG. \ref{KvsPi}(b), the parameters with errors are
\be 
\begin{aligned}
\alpha_0&= 0.00616_{\pm 1.56\times 10^{-8}},
\quad \alpha_2 = 0.00690_{\pm 4.38\times 10^{-5}},\\
\alpha_3 &= 0.00518_{\pm 1.05\times 10^{-3}},\quad \alpha_4 = 0.447_{\pm 6.50\times 10^{-3}}.
\end{aligned}
\ee 

Eq.\eqref{a012} is understood as the constraint equation for $K_f,\pi_f$ supported by the numerical results. Comparing \eqref{a012} to the Friedmann equation, we find that the right-hand side of \eqref{a012} gives an effective scalar density $\rho_{\rm eff}$, which contains not only a $\pi_f^2$ term but also higher derivative terms with $\pi_f^3$ and $\pi_f^4$. 

The effective scalar density also contains $\alpha_0$, which is understood as an effective scalar potential due to its dependence on $\phi_f$. $\alpha_0>0$ plays a role similar to an effective positive cosmological constant. The nonzero $\alpha_0$ indicates that on the final slice $K_{\rm crit} > K_i$, implying the universe is acceleratedly expanding. 

\begin{figure}[h]
\bigskip
    \centering \includegraphics[scale=0.062]{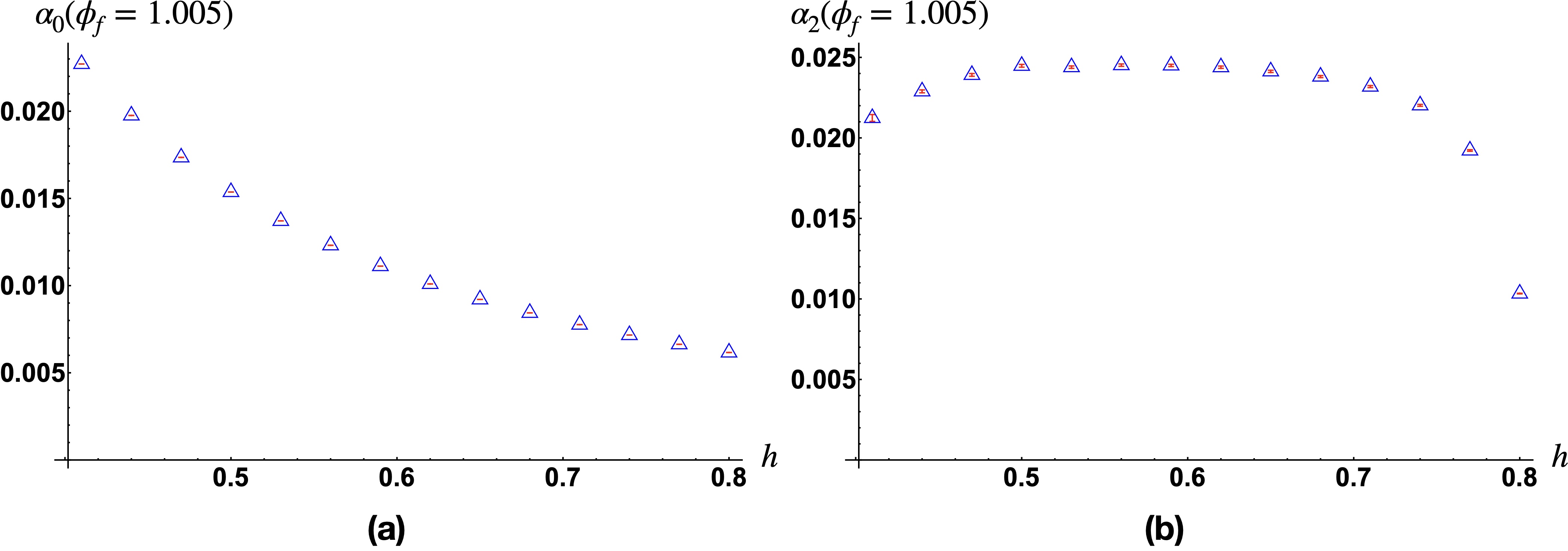}
    \caption{The relations between $\alpha_0, \alpha_2$ and $h$ at $\phi_f=1.005$. The blue uptriangles represent the values of $\alpha_0$ and $\alpha_2$ as they vary with $h$. The error bars are denoted by the red fences.}
    \label{a0a2Vsh}
\end{figure}

To understand the dependence of $\alpha_0,\alpha_2,\alpha_3,\alpha_4$ on $\phi_f$ and $h$, we vary both $\phi_f$ and $h$ and perform the same algorithm for computing $\alpha_0,\alpha_2,\alpha_3,\alpha_4$. As a result, FIG.\ref{ParamsvsPhi2} shows the relation between $\alpha_0,\alpha_2,\alpha_3,\alpha_4$ and $\phi_f$ at a fixed $h=0.8$. Particularly, the fact that $\alpha_0$ depends on $\phi_f$ suggests it to be an effective scalar potential. Since $\alpha_2$ relates to the effective gravitational constant, the property that $\alpha_2$ depends on $\phi_f$ seems to indicate that the effective theory from the spinfoam coupled to scalar matter relates to the non-minimal coupling between gravity and scalar matter\footnote{An example of non-minimal coupling between gravity and scalar matter is given by the term $\frac{1}{16\pi G}\varphi R$, which couples the scalar matter to the scalar curvature in the Lagrangian. This term can be re-written as $\frac{1}{16\pi G(\varphi)} R$, resembling the usual Einstein-Hilbert term, but with the gravitational constant $G(\varphi)=G\varphi^{-1}$ becoming $\varphi$-dependent.}. FIG.\ref{a0a2Vsh} shows the relation between $\alpha_0, \alpha_2$ and $h$ at $\phi_f=1.005$.


\subsection{Double-hypercube Model}

In the double-hypercube model, ensuring the nondegeneracy of the action at both real and complex critical points within $\mathcal{Z}_{\mathcal{K}}$ involves setting $N_{\rm out}=8$ in (\ref{Z1}). The eight external spins are defined as:
\begin{equation}
\begin{split}
    j^{\text{out}}_I=&\{j_{1,9,14},j_{1,11,16}, j_{1,2,16},j_{1,10,16},\\
    &j_{1,9',14'},j_{1,11',16'}, j_{1,2,16'},j_{1,10',16'}\},
\end{split}\label{joutdouble}
\end{equation}
The relation between $\{j_{1,9,14},j_{1,11,16}, j_{1,2,16},j_{1,10,16}\}$ and $h_i$ is given by (\ref{jandh}) with the substitutions $h\to h_i$ and $a_f\to a_m$. Similarly, the relation between $\{j_{1,9',14'},j_{1,11',16'}, j_{1,2,16'},j_{1,10',16'}\}$ and $h_f$ is (\ref{jandh}) with the substitutions $a_i\to a_m$ and $h\to h_f$.

The scalars on the double-hypercube are $\varphi_{v}$ with $v=1,2,\ldots, 48$. We consider the length variables $(a_i, a_f, L_i, L_f, a_m)$ (See FIG.\ref{Oneand2Hypercube} (b) ) entering the scalar action. We have to change the spin variables $(j_{1,2,8}, j_{9,14,16}, j_{9',14',16'},j_{1,4,16},j_{1,4,16'})$ in the spinfoam to the length variables $(\frak{a}_i,\frak{a}_f,\frak{l}_i,\frak{l}_f, \frak{a}_m)$ to couple the spinfoam to scalar matter. The change of variables is given explicitly by: 
\be
\begin{aligned}
    j_{1,2,8} = \frac{\frak{a}_m^2}{\sqrt{2}}, \quad
    j_{9,14,16} = \frac{\frak{a}^2_i}{\sqrt{2}}, \quad 
    j_{9',14',16'} = \frac{\frak{a}^2_f}{\sqrt{2}},\\
    j_{1,4,16(16')} = \frak{a}_{m(f)} \sqrt{\frac{\frak{l}_{i(f)}^2}{2}-\left(\frac{\frak{a}_{i(f)}+\frak{a}_m}{2}\right)^2}.\label{jjjjj}
\end{aligned}
\ee
The lengths $\fa_i,\fa_f,\fl_i,\fl_f,\fa_m>0$ can be uniquely recovered by $j$'s. Geometrically $({\sqrt{\gamma}}\frak{a}_i,{\sqrt{\gamma}}\frak{a}_f,{\sqrt{\gamma}}\frak{l}_i,{\sqrt{\gamma}}\frak{l}_f,{\sqrt{\gamma}}\frak{a}_m)$ corresponds to the lengths $({a_i}, {a_f}, {L}_i,L_f,a_m)$. However, the same as before, $\fa_i,\fa_f$ are the integration variables changed from the above $j$'s, and they are distinguished from the external parameters $a_i,a_f$ that enter the coherent state data $j_b^0$ and external spins $j_I^{\rm out}$.

\begin{figure}[h]
    \centering
    \includegraphics[scale=0.06]{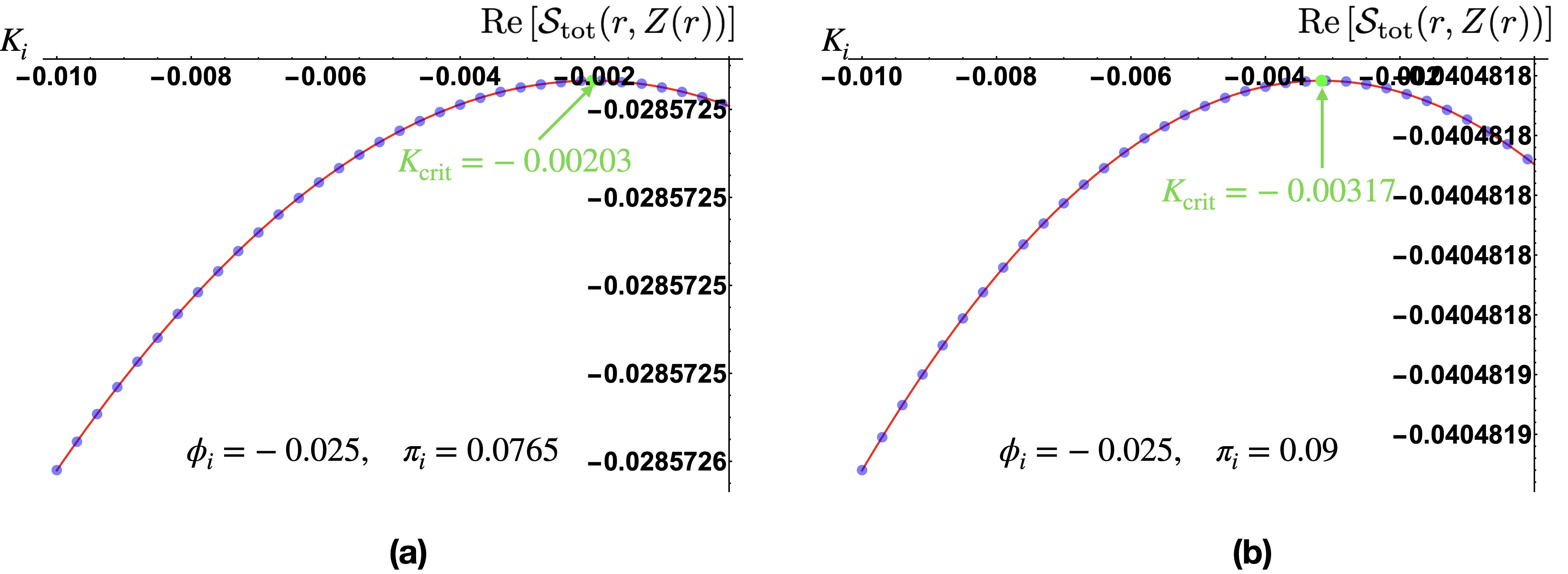}
    \caption{The real part of the total action $\mathcal{S}_{\text{tot}}(r,Z(r))$ at the complex critical point $Z(r)$ versus the extrinsic curvature $K_i$ at the initial slice $t=-h=-0.8$ with the boundary conditions (a) $\phi_i=-0.025, \pi_i=0.0765$ and (b) $\phi_i=-0.025, \pi_i=0.09$. The red curves are approximate functions that interpolate the data of the blue points. The green points are the location $K_{\rm crit}$ of the maxima as determined by the interpolation of the function.}
    \label{ReSVsK2Hyp}
\end{figure} 
\begin{figure}[h]
    \centering
    \bigskip
    \includegraphics[scale=0.12]{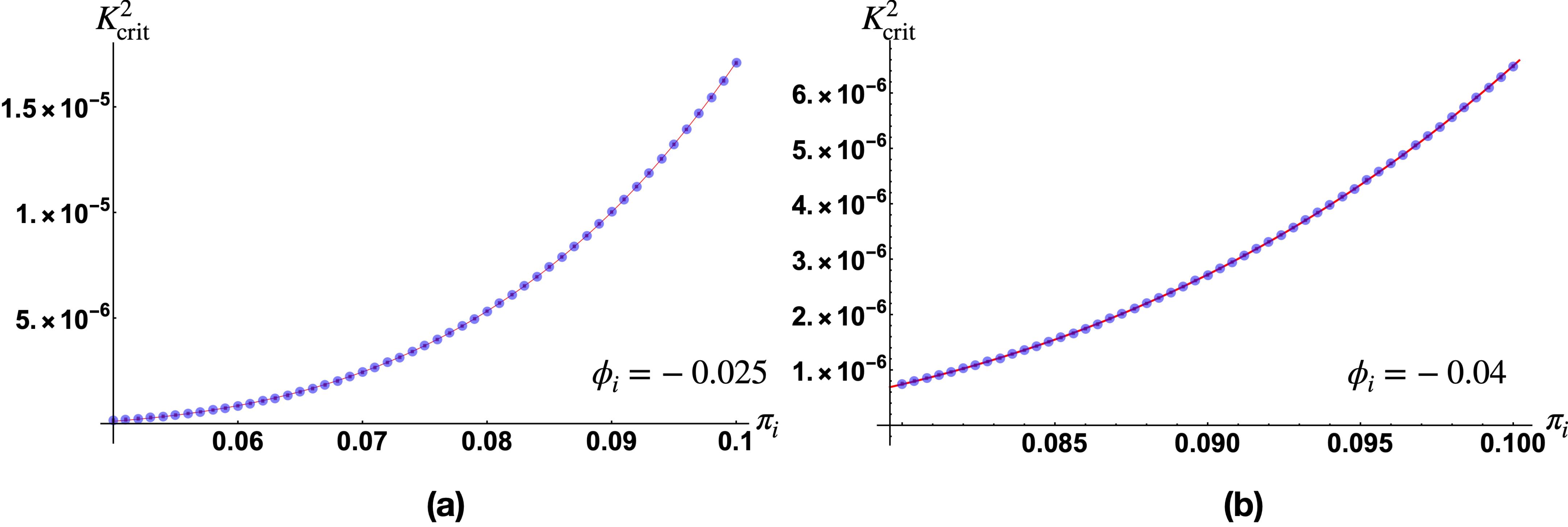}
    \caption{The saddle point $K^2_{\rm crit}$ changes with varying $\pi_i$ at different $\phi_i=-0.025$ in (a) and $\phi_i=-0.04$ in (b). The red curves represent the polynomial fits with \eqref{a012Hypers}.}
    \label{KvsPi2Hyper}
\end{figure}

The amplitude $\mathcal{Z}_{\mathcal{K}}\left(j^{\text{out}}_I\right)$ involves $N=2354$ dimensional real integrals, similar to (\ref{Z2}), with the Jacobian  $\mathbf{J}_{\mathbf{f}}(\frak{a}_i,\frak{a}_f,\frak{l}_i,\frak{l}_f, \frak{a}_m)$. The integrated variables, subject to the gauge fixing and parametrization discussed in Appendix \ref{Gaugefix},  are as follows:
\be 
\mathbf{x}=\{g_{ve}, z_{vf},  \xi^{\pm}_{eh},l_{eh}^+, j_{\bar{h}}, \frak{a}_i,\frak{a}_f,\frak{l}_i,\fl_f,\fa_m, \varphi_v\}. 
\ee
Here, $j_{\bar h}$ denotes spins other than the ones in \eqref{joutdouble} and \eqref{jjjjj}.


To make an analog of the (time-reversal) symmetic cosmic bounce,   we impose the following conditions for the initial and final data: 
\be 
a_i=a_f=a,\quad K_f=-K_i>0\\
\phi_f=-\phi_i>0, \quad \pi_f=\pi_i>0. 
\ee 
The property that $K\propto \dot{a}$ is negative at the initial and positive at the final indicates contracting and expanding universes at the initial and final slices. The fact that $\dot{a}$ evolves from negative to positive indicates a cosmic bounce occurring in the evolution.


We fix the other external data $j_I^{\rm out}$ by setting $h_f=h_i=h=0.8$, and we identify the real critical point corresponding to the flat geometry with $a=1=a_m$, $K_i=\phi_i=\pi_i=0$. When deforming $K_i,\phi_i,\pi_i$ to nonvanishing values, we perform numerical computations to determine the solution of the complex critical equations in the same way as in the single-hypercube model. Following the same algorithm and  fixing $a=1$, for any values of $h_i,h_f,\phi_i,\pi_i$, we perform the computation of the complex critical point for a range of $K_i$ values. An interpolation for $\Re[\mathcal{S}_{\text{tot}}(r,Z(r))]$ as a function of $K_i<0$ enables us to identify the location $K_{\rm crit}$ of the maximum, as illustrated in FIG. (\ref{ReSVsK2Hyp}). Recall that the maxima of $\Re[\mathcal{S}_{\text{tot}}(r,Z(r))]$ and $|\cz_\ck|$ are approximately located at the same point. We suggest that the spinfoam amplitude with nonzero $K_i=-K_f=K_{\rm crit}$ should have the interpretation as the quantum analog of a cosmic bounce.


\begin{figure}[h]
    \centering
    \includegraphics[scale=0.115]{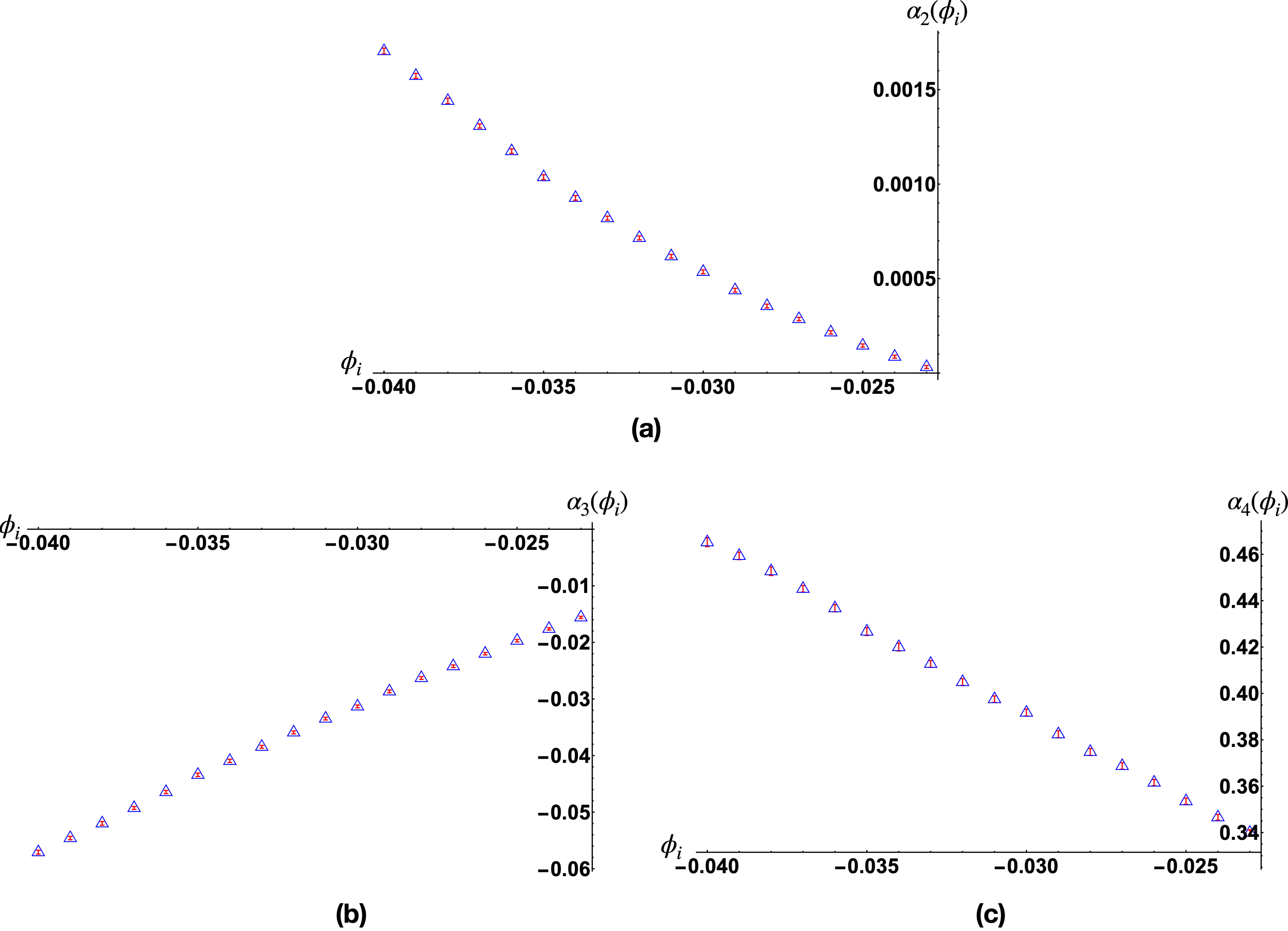}
    \caption{(a), (b) and (c) show the relation between $K_{\rm crit}^2$ and $\phi_i$ using the model given by (\ref{a012Hypers}). The blue uptriangles denote $\alpha_2$ in (a), $\alpha_3$ in (b), and $\alpha_4$ in (c) as they change with $\phi_i$. 
    }
    \label{a0123hypers}
\end{figure}
\bigskip
\begin{figure}[h]
    \centering
    \bigskip\bigskip
    \includegraphics[scale=0.08]{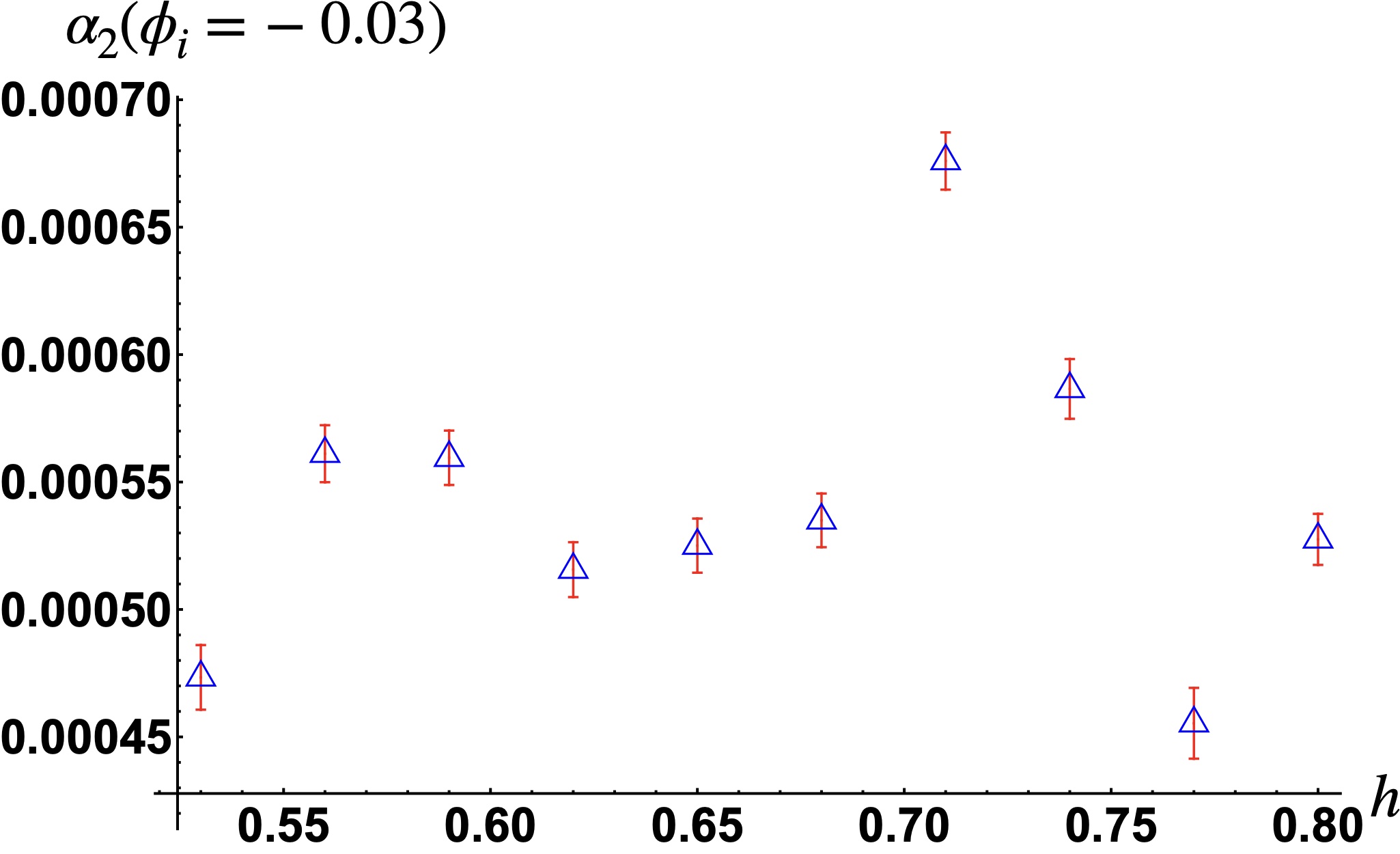}
    \caption{shows the relationship between $\alpha_2$ and $h$ using the algorithm we describe above at $\phi_i=-0.03$.}
    \label{alpha2vsh}
\end{figure}

The relation among $K_{\rm crit}$ and other boundary data gives a constraint comparable to the Friedmann equation. FIG. \ref{KvsPi2Hyper} provides examples of numerical data indicating the relation between $K_{\rm crit}$ and $\pi_i$ with other external parameters fixed. The relationship between $K^2_{\rm crit}$ and $\pi_i$ can be accurately modeled by a polynomial expression of the form: 
\be 
K^2_{\rm crit} = \a_2(\phi_i)\pi^2_i+\a_3(\phi_i)\pi^3_i+\a_4(\phi_i)\pi^4_i+O(\pi^5_i).\label{a012Hypers}
\ee
The coefficients $\a_2, \a_3, \a_4$ can be determined by polynomial fit. $K_{\rm crit}=0, \phi_i=0, \pi_i=0$ result in the real critical points (with $\mathrm{Re}(\cs_{\rm tot})$=0) corresponding to flat geometry without scalar. It has already been predicted that the constant term $\alpha_0$ should vanish when $\phi_i=0$. The numerical results suggest that $\alpha_0$ should still be negligible when $\phi_i\neq 0$. The fitting model \eqref{a012Hypers} without $\a_0$ and $\alpha_1$ gives smaller error bars. Specifically, for the case where $\phi_i=-0.025$ in FIG. \ref{KvsPi2Hyper}(a), the coefficients with errors are:
\begin{equation}
\begin{split}
\a_2= 0.000146_{\pm 8.40\times 10^{-6}},\qquad\qquad\\
\a_3= -0.0197_{\pm 2.14\times 10^{-4}},\quad
\a_4 = 0.354_{\pm 0.00133}. 
\end{split}
\end{equation}
For the scenario with $\phi_i=-0.04$ in FIG. \ref{KvsPi}(b),
\be 
\begin{aligned}
\a_2 = 0.00170_{\pm 1.56\times10^{-5}},\qquad\qquad\\
\a_3 = -0.0571_{\pm 3.44\times10^{-4}},\quad 
\a_4 = 0.465_{\pm 0.00190}.
\end{aligned}
\ee 
We exclude the linear term in (\ref{a012Hypers}) because adding the linear term produces larger errors in the polynomial fit, similar to the situation in \eqref{a012}.


We suggest that Eq.\eqref{a012Hypers} should be understood as a modified Friedmann equation when a symmetric bounce happens.\footnote{The spinfoam dynamics prevents the formation of curvature singularities, as suggested in \cite{Rovelli2013d}. Hence the bouncing behavior here can be genuily intepreted as a process without the classical big bang singularity.} \, The right-hand side of \eqref{a012Hypers} should be proportional to the effective scalar density $\rho_{\rm eff}$, which contains higher derivative terms with $\pi_i^3$ and $\pi_i^4$. The scalar potential vanishes since the constant term vanishes in \eqref{a012Hypers}, as a consequence of the symmetric bounce. 

Finally, FIG.\ref{a0123hypers} plots the dependence of $\alpha_2,\alpha_3,\alpha_4$ on $\phi$, and FIG.\ref{alpha2vsh} plots the dependence of $\alpha_2$ on $h$. Similar to the single-hypercube, the property that $\a_2$ depends on $\phi$ suggests a non-minimal coupling between gravity and scalar in the effective theory.

\pagebreak

\section{Conclusion and Outlook}

We have applied the numerical method of complex critical points to the 4d Lorentzian EPRL spinfoam amplitude on two different simplicial complexes: the single-hypercube complex and double-hypercube complex. These are designed to approximately model spatial homogeneous and isotropic cosmology. The coupling of  scalar matter to spinfoams yields nontrivial physical implications for the effective cosmological dynamics. In the single-hypercube model, the numerical results suggest an effective Friedmann equation with an effective scalar density containing higher-order derivative terms. The presence of a non-zero effective scalar potential implies the presence of an effective positive cosmological constant, leading to an accelerated expansion of the universe. In the double-hypercube model, we explore a scenario analog to a symmetric cosmic bounce. We obtain a similar effective Friedmann equation with an effective scalar density, which still contains higher-order derivative terms but has negligible scalar potential.


This work not only contributes to the understanding of quantum cosmology within the full covariant LQG framework, but also extends the exploration of matter coupling in the spinfoam formalism. Our results open avenues for further investigations; we list some possible directions. 

Firstly, the double-hypercube model seems to allow the bounce to happen at low density (with small $\pi_i$), so it is more like an analog to the $\mu_0$-scheme LQC. This may be the consequence of only considering two hypercubes and not taking into account the possible change of (spatial) lattice in the time evolution. Similar phenomena happens in the derivation of LQC from the full canonical LQG \cite{Han:2019vpw,Dapor:2017rwv}. It is suggested in \cite{Han:2021cwb} that allowing the lattice to be dynamical should give a dymanics similar to that of the $\bar{\mu}$-scheme in LQC, in which the bounce only happens at the critical energy density, taken to be Planckian. A dynamical lattice means that the number of cubes is changing at different steps in the evolution.  

Secondly, the current model only involves two hypercubes in the time evolution. To enhance the model's realism and to test the reliability of the truncation, we can increase degrees of freedom by incorporating more hypercubes along the $t$-direction, leading to a more accurate representation of physical systems.

Finally, we find $K_{\rm crit}$ by locating the maximum of the amplitude's absolute value within a subspace of parameters, i.e., with fixed $a_{i(f)},\phi_{i(f)},\pi_{i(f)}$. This can be  improved by  scanning the entire space of boundary data and finding the maximum of the amplitude's absolute value. Another desirable improvement could be to develop techniques for fitting the constraint among $(a_{i(f)}, K_{i(f)},\phi_{i(f)},\pi_{i(f)})$ within a higher-dimensional parameter space. Performing statistical analyses to assess the robustness and reliability of spinfoam predictions in a higher-dimensional parameter space. Simultaneously, investigating the stability of the results will enhance the predictive power of the spinfoam formalism.

\bigskip
\begin{acknowledgments}
The authors acknowledge Bianca Dittrich and Carlo Rovelli for helpful discussions. This work benefits from the visitor's supports from Beijing Normal University, FAU Erlangen-N\"urnberg, the University of Western Ontario, and Perimeter Institute for Theoretical Physics. MH receives support from the National Science Foundation through grants PHY-2207763, the College of Science Research Fellowship at Florida Atlantic University, and a visiting professorship at FAU Erlangen-N\"urnberg. MH, HL and DQ are supported by research grants provided by the Blaumann Foundation. 
FV’s research at Western University is supported by the Canada Research Chairs Program, the Natural Science and Engineering Council of Canada
(NSERC) through the Discovery Grant ``Loop Quantum Gravity: from Computation to Phenomenology'', and the John Templeton Foundation trough the ID\# 62312 grant ``The Quantum Information Structure of Spacetime'' (QISS). FV's research at the Perimeter Institute is supported through its affliation program.
The Government of Canada supports research at Perimeter Institute through
Industry Canada and by the Province of Ontario through the Ministry of Economic Development and Innovation.
Western University and Perimeter Institute are located in the traditional lands of Anishinaabek, Haudenosaunee, L\=unaap\`eewak, Attawandaron, and Neutral peoples.
\end{acknowledgments}

\bibliographystyle{jhep}
\bibliography{ref}

\pagebreak
\onecolumngrid
\appendix

\section{Gauge freedom and gauge fixing in spinfoam action} \label{Gaugefix}

Since the spinfoam amplitude expressed in \eqref{SFamplitude} includes five types of continuous gauge degrees of freedom, it is necessary to introduce gauge fixings to eliminate the gauge degrees of freedom.
\bi
\item There is the $\Slc$ gauge transformation at each $v$:  
\be 
g_{ve}\mapsto x_v^{-1}g_{ve},\quad z_{vf}\mapsto x_v^{\rm T}z_{vf}, \quad x_v\in\Slc. 
\ee
To fix this gauge degree of freedom, we select one $g_{ve}$ to be a constant $\Slc$ matrix for each 4-simplex. The amplitude is unaffected by the choice of constant matrices.

\item For each $z_{vf}$, there is the scaling gauge freedom:
\be 
z_{vf}\mapsto\lambda_{vf} z_{vf}, \qquad \lambda_{vf}\in\mathbb{C}.  \label{zvf}
\ee 
We fix the gauge by setting one of the components of $z_{vf}$ to 1, i.e. either $z_{vf}=\left(1, {\a}_{vf}\right)^\mathrm{T}$ or $z_{vf}=\left({\a}_{vf},1\right)^\mathrm{T}$, where ${\a}_{vf}\in\C$. We adopt $z_{vf}=\left({\a}_{vf},1\right)^\mathrm{T}$, when the first component of $\mathring{z}_{vf}$ 
happens to be zero at the critical point.

\item At each spacelike triangle $h$ in a timelike tetrahedron (all timelike tetrahedra are internal), 
\be
\xi_{eh}^\pm\mapsto e^{i\psi_{eh}^\pm} \xi_{eh}^\pm,\qquad \psi^\pm_{eh}\in\mathbb{R},
\ee
leaves the action $S$ invariant. We fix the gauge by parametrizing 
\be
\xi^+_{eh}= \left(\cosh\theta_{eh},e^{-i\b_{eh}}\sinh\theta_{eh}\right)^{\rm T}, \quad \xi^-_{eh}= \left(e^{i\b_{eh}}\sinh\theta_{eh},\cosh\theta_{eh}\right)^{\rm T},\qquad \beta_{eh},\theta_{eh}\in\mathbb{R}.\label{xipm}
\ee 

\item At each timelike triangle $h$ (all timelike triangles are internal), the action $S$ is invariant under the scaling for $l^+_{eh}$
\be
l^+_{eh}\mapsto\lambda_{eh} l^+_{eh},\qquad \lambda_{eh}\in\mathbb{C}.
\ee
The scaling cancels between $F_{veh}$ and $F_{v'eh}$ for the orientation of $\partial h^*$ is outgoing from the vertex $v^*$ and incoming to $v'^*$, so it leaves the action $S$ invariant. Rigorously speaking, this scaling symmetry is not a gauge freedom if we parametrize $l^+_{eh}$ by $l^+_{eh}=v_{eh}\cdot (1,1)^{\rm T}$ with $v_{eh}\in \mathrm{SU}(1,1)$. But we still can use this symmetry to simplify the parametrization of $l^+_{eh}$. Namely, for any $l^+_{eh}=(l^{0+}_{eh},l^{1+}_{eh})^{\rm T}=v_{eh}\cdot (1,1)^{\rm T}$, we can modify $l^+_{eh}$ by $l^+_{eh}=(1,l^{1+}_{eh}/l^{0+}_{eh})^{\rm T} $ where $l^{1+}_{eh}/l^{0+}_{eh}\equiv e^{i\zeta_{eh}}$ is a phase implied by $v_{eh}\in \mathrm{SU}(1,1)$. Therefore we parametrize
\be
l^+_{eh}=(1,e^{i\zeta_{eh}})^{\rm T}
\ee
in the action, and $l_{eh}^-$ satisfies $\langle l^+_{eh},l^-_{eh}\rangle =1$ and $\langle l^-_{eh},l^-_{eh}\rangle =0$.

\item There is the $\Su$ gauge transformation for each bulk spacelike tetrahedron  $e$: 
\be 
g_{v'e}\mapsto g_{v'e}h_e^{\rm T},\quad g_{ve}\mapsto g_{ve}h_e^{\rm T}, \quad h_e\in\Su. \label{gaugesu2}
\ee 
To fix this $\Su$ gauge freedom, we parameterize either $g_{ve}$ or $g_{v'e}$ by a lower triangular matrix of the form
\begin{equation}
	k'=\left(\begin{matrix}
		\lambda^{-1}&0\\\mu &\lambda
	\end{matrix}\right),\ \lambda\in\mathbb{R}\setminus\{0\},\ \mu\in\mathbb{C} .\label{upper}
\end{equation} 
For any $g\in\Slc$, we can write $g=(g_0^\dagger)^{-1}$ and use the standard Iwasawa decomposition $g_0=ku$ where $k$ is an upper triangular matrix. Then we obtain $g=k'u$ where $k'=(k^\dagger)^{-1}$ is lower triangular. Choosing $h_e^{\rm T}=u^{-1}$ transforms $g_{v'e}$ to the lower triangular matrix $k'$.


\item Every timelike tetrahedron in our model contains at least one spacelike triangle and one timelike triangle, and all timelike tetrahedra are internal. There is the $\mathrm{SU}(1,1)$ gauge transformation for each timelike tetrahedron $e$. 
\be 
g_{v'e}\mapsto g_{v'e}h_e^{\rm T},\quad g_{ve}\mapsto g_{ve}h_e^{\rm T}, \quad \xi_{eh}^\pm\mapsto h_e\xi_{eh}^\pm,\quad l_{eh}^\pm \mapsto  h_e l_{eh}^\pm,\quad 
\qquad h_e\in\mathrm{SU}(1,1). \label{gaugesu11}
\ee  
We implement the following procedure to fix this gauge freedom:  Firstly, We choose a spacelike triangle $f$ and find a $\mathrm{SU}(1,1)$ gauge transformation $U_e$ such that 
 \be
	\xi^+_{ef}, \xi^-_{ef} \xRightarrow[]{U_{e}\in\mathrm{SU}(1,1)} \left(\begin{matrix}
		1\\
		0
	\end{matrix}\right), \,\left(\begin{matrix}
		0\\
		1
	\end{matrix}\right). 
	\ee 
The corresponding $U_e$ acts on all $\xi^{\pm}_{ef}$ or $l^\pm_{eh}$ in this tetrahedron. Secondly, choose a timelike triangle $h$ and rewrite the corresponding $l^+_{eh}=(e^{-i\zeta_{eh}/2},e^{i\zeta_{eh}/2})^{\rm T}$ by the scaling symmetry.  
Then we make a futher $\mathrm{SU}(1,1)$ gauge transformation 
	\be
	\tilde{U}_e= \left(\begin{matrix}
		e^{i\zeta_{eh}/2} & 0\\
		0& e^{-i\zeta_{eh}/2}
	\end{matrix}\right). 
	\ee 
This matrix $\tilde{U}_e$ fixes $l^+_{eh}$ to $(1,1)^{\rm T}$. 
This $\tilde{U}_e$ again acts on all $\xi_{ef}^\pm$ and $l_{eh}^\pm$ within the same tetrahedron. In particular, for the spacelike $f$, $\xi_{ef}^\pm$ become
 $	 \xi^+_{ef}= (	e^{i{\zeta_{eh}}/{2}},0)^{\rm T}, \   \xi^-_{ef} = (0, e^{-i{\zeta_{eh}}/{2}})^{\rm T}$, whereas the phases can be further removed by the gauge transformation of $\xi_{ef}^\pm$. These gauge transformations allow us to gauge fixing the timelike face $l^\pm_{eh}$ are in the form: 
\be
l^+_{eh}= \left(\begin{matrix}
		1\\1
	\end{matrix}\right), \qquad 
\xi^+_{ef}= \left(\begin{matrix}
		1\\
		0
	\end{matrix}\right), \qquad
\xi^-_{ef}=\left(\begin{matrix}
		0\\
		1
	\end{matrix}\right)\label{GFform}
	\ee 
 
\end{itemize}

For a generic data $(g_{ve},z_{vf},\xi_{eh}^{\pm},l^+_{eh})$, firstly, we can use the SU(1,1) gauge transformation to fix $\xi_{ef}^{\pm}$ and $l_{eh}^+$ into the gauge fixing form \eqref{GFform} for all timelike internal tetrahedra. Secondly, we use the $\mathrm{SL}(2,\C)$ gauge transformation to fix one $g_{ve} $ to be constant within each 4-simplex. In our model, every 4-simplex has at least one timelike tetrahedron. We always choose the gauge-fixed $g_{ve}$ to associate with the timelike $e$, if $v$ does not have a boundary tetrahedron, or with the boundary (spacelike) $e$, if $v$ has the boundary tetrahedron. Since the $\mathrm{SL}(2,\C)$ gauge transformation acts to the left of $g_{ve}$, it does not affect the SU(1,1) gauge fixing \eqref{GFform}. Thirdly, we use the SU(2) gauge transformation to put one of $g_{ve}$ into a lower triangular matrix for each spacelike internal $e$. The SU(2) gauge transformation does not affect the previous gauge fixings since it only acts on spacelike tetrahedra and acts to the right of $g_{ve}$. Lastly, we use the scaling symmetry to reduce $z_{vf}, \xi^{\pm}_{eh},l^+_{eh}$ to the gauge-fixing forms. This procedure can transform any data $(g_{ve},z_{vf},\xi_{eh}^{\pm},l^+_{eh})$ to the gauge-fixing form defined above, so it shows the above gauge fixing is well-defined.

\section{Coherent states in the gravitational field and the scalar field}\label{Coherent states in the gravitational field and the scalar field}
Consider the Friedmann–Lema\^itre–Robertson–Walker (FLRW) cosmology with $k=0$ coupled to a massless scalar field. To avoid integration divergence, let us choose a fiducial cell $\mathcal V_{\mathbb{R}^3}$ and restrict the integrals over the fiducial cells. The volume of the fiducial cell under the fiducial metric $\delta_{ab}$ will be denoted by $V_o$. The FLRW metric is
\be\label{eq:cosmologyMetric}
ds^2=-\mathrm{d}t^2 + \frac{a(t)^2}{V_o^{2/3}}\left(\mathrm{d}x^2+\mathrm{d}y^2+\mathrm{d}z^2\right), \label{dsFLRW}
\ee 
where $t$ represents cosmic time, $a(t)$ denotes the scale factor, and the rescaling factor $1/V_o^{2/3}$ is introduced to cancel out the volume of the fiducial cell $\int_{\mathcal{V}} d^3x = V_o$. Here, we will present coherent states in both the gravitational field and the scalar field with this metric.

Substituting the metric \eqref{eq:cosmologyMetric} into the Einstein-Hilbert action, we obtain 
\be 
S_{\rm GR} = \frac{1}{16\pi G}\int_{\mathbb R\times \mathcal V}  \sqrt{-g} R\mathrm{d}^4x = -\frac{3}{8\pi G}\int_{\mathbb R}a(t)\dot{a}(t)^2 \mathrm{d} t,
\ee
where we apply $\sqrt{-g}=V_o^{-1}a^3$ and $R = \frac{6(\dot{a}^2 + a\Ddot{a})}{a^2}$ and ignore the boundary term. The expression of the action implies the canonical variable
\begin{equation}\label{eq:pac}
p=a^2,\quad c=\dot a=\frac{\dot p}{2\sqrt p},
\end{equation}
 satisfying the commutation relation 
\begin{equation}\label{eq:possoncp}
\{c,p\}=\frac{8\pi G}{3}.
\end{equation}
Given an elementary face $\mathcal A$ in the fiducial cell $\mathcal V$, it can be verified that $p$ is equal to the area of $\mathcal A$. In other words, $A=p/(8\pi \gamma G\hbar)$ shares the same geometric interpretation as the variable $j$ used in \eqref{psiDef}, i.e., the area of a surface with respect to the Planck area $8\pi \gamma G\hbar$. 

To investigate the geometric meaning of the parameter $\vartheta_0$ in \eqref{psiDef}, let us consider the coherent state
\be \label{eq:coherentstate}
\psi_{(A_0,\vartheta_0)}(A) =\mathcal N \exp\left[-\frac{1}{2}\frac{(A - A_0)^2}{A_0} - i \gamma A\vartheta_0\right],
\ee 
in the Sch\"odinger representation, where $\mathcal N$ denotes the normalization factor. Due to the Poisson bracket \eqref{eq:possoncp}, the variable $c$ should be promoted as the operator
\begin{equation}
\hat c=\frac{i}{3\gamma} \frac{\mathrm d}{\mathrm d A}.
\end{equation} 
The expectation value of $\hat c$ with respect to the coherent state \eqref{eq:coherentstate} reads
\begin{equation}
\langle \hat c\rangle=\frac{1}{3}\vartheta_0.
\end{equation}
Taking into account the classical relation \eqref{eq:pac} between $c$ and $\dot a$, we get
\begin{equation}
\vartheta_0=3\dot a = K a, 
\end{equation} 
where $K$ denotes the trace of the extrinsic curvature of the spatial slice $t=\text{const}$. It iwll be seen that this result coincide with the result given in Appendix \ref{Kandtheta}.

For the homogeneous scalar field $\phi(t)$, its action is
\be 
S_{\mathrm{scalar}} = -\frac{1}{2}\int_{\mathbb R\times\mathcal V}  \sqrt{-g}\left(g^{ab}\nabla_a\phi \nabla_b\phi\right)\dd^4 x= \frac{1}{2}\int_{\mathbb R} a(t)^3 \dot \phi(t)^2\dd t,
\ee 
where we use $g^{ab} \nabla_a\phi \nabla_b\phi = -\dot\phi^2$. This action indicates the canonical pairs $\phi$ and $\pi_\phi=a^3\dot\phi$ satisfying the Poisson bracket
\begin{equation}\label{eq:poissonscalar}
\{\phi,\pi_\phi\}=1.
\end{equation}

The coherent state for the scalar field  is given by
\be 
\psi_{z}(\phi) = \exp\left[\frac{1}{\hbar}\left(\frac{z^2_0}{4}-\frac{(\phi-z_0)^2}{2}-\frac{z_0\bar{z}_0}{4}\right)\right].
\ee 
Here, $z_0$ and $\bar{z}_0$ provide the complex parametrization of the scalar sector in the phase space:
\be 
z_0= \phi_0+\imath \pi_0, \quad \bar{z}_0=\phi_0-i \pi_0.
\ee
The Poisson bracket \eqref{eq:poissonscalar} lead us to quantize $\pi_\phi$ as
\begin{equation}
\hat\pi_\phi=-i\hbar\frac{\dd}{\dd\phi}. 
\end{equation}
Then, we get the expectation values of observables $\hat{\pi}$ and $\hat{\phi}$  as
\be 
\langle \hat{\pi}\rangle=\frac{z_0-\bar z_0}{2i}=\pi_0, \quad \langle \hat{\phi}\rangle =\frac{z_0+\bar z_0}{2}=\phi_0, 
\ee  
relating the parameters $z_0$ and $\bar z_0$ to the expectation values of observables. 

\section{Derivation of relation between Extrinsic curvature and dehedral angles} \label{Kandtheta}
The generic spatially flat FLRW metric is given by
\be 
ds^{2}=-dt^{2}+a^{2}(dx^{2}+dy^{2}+dz^{2}), \label{FLRW metric}
\ee 
where $t$ is the proper time, $a$ denotes the scale factor of the universe. Eq. \eqref{FLRW metric} corresponds to $V_o=1$ in Eq.(\ref{dsFLRW}). The extrinsic curvature at any $t$ is given by
\be 
K_{ab} = \frac{1}{2} \dot{h}_{a b} = \frac{1}{2} \partial_t (a^2) \delta_{ab}= a \dot{a}\delta_{ab}, \label{E2}
\ee 
where we use the relation between $\dot{h}_{ab}$ and $K_{ab}$ that \cite{Wald:1984rg}
\be 
\dot{h}_{a b}=2 N K_{a b}+\mathscr{L}_{\vec{N}} h_{a b} = 2K_{ab},
\ee 
by $N=1$ and $\mathscr{L}_{\vec{N}} h_{a b}=0$. $K$ is the trace of the extrinsic curvature and is given by
\be 
K=h^{ab}K_{ab}=3a^{-2}a\dot{a}=\frac{3\dot{a}}{a}, 
\ee 
The discretization of the Gibbons-Hawking boundary term in Regge calculus \cite{Hartle:1981cf,1988CQGra...5.1193B} is
\be 
\mathrm{Ar}_b \vartheta^{0}_{b} = \int \mathrm{d}^3 x \sqrt{h} K = V_{A_f} K.
\ee 
Here, $\vartheta^{0}_{b}$ is the boundary dihedral angle at the boundary triangle $b$, and $\mathrm{Ar}_b$ is the area of $b$, $V_{A_f}$ represents a volume associated to $b$ (see FIG \ref{ratio}(a)). Therefore, the boundary dihedral angle $\vartheta^{0}_{b}$ related to the simplicial extrinsic curvature $K$ is given by
\be 
\vartheta^{0}_{b} = K\times \frac{V_{A_f}}{A_f}.
\ee

\section{Triangulation of 4D Hypercube and the periodic boundary conditions} \label{1Hypercube}
In 4D, a hypercube has 16 vertices, 32 edges, 24 faces, and 8 3D cubes. These cubes are classified by their normal directions as follows: 
\be
 \begin{aligned}
t-\text{direction cubes:}\quad 	& (1,2,3,4,5,6,7,8),\,(9, 10, 11, 12, 13, 14, 15, 16),\\
x-\text{direction cubes:}\quad 	& (1, 2, 3, 4, 9, 10, 11, 12),\, (5, 6, 7, 8, 13, 14, 15, 16),\\
y-\text{direction cubes:}\quad 	& (1, 2, 5, 6, 9, 10, 13, 14)\,(3, 4, 7, 8, 11, 12, 15, 16), \\
z-\text{direction cubes:}\quad 	& (1, 3, 5, 7, 9, 11, 13, 15),\,(2, 4, 6, 8, 10, 12, 14, 16).
 \end{aligned}
\ee 
These 8 cubes are the starting point for creating the 4-simplices. The hypercube we choose is a convex hull of all points whose Cartesian coordinates $(t, x, y, z)$ are 
\be
\left(0\,\text{or } h, 0\,\text{or } a, 0\,\text{or } a, 0\,\text{or } a\right)\quad \text{with }h>0, a>0. \label{coordApp}
\ee 
Here, the cube $(1,2,3,4,5,6,7,8)$ is located on the slice $t=0$, representing the past cube, while the cube $(9, 10, 11, 12, 13, 14, 15, 16)$ is on the slice $t=h$, representing the future cube.

\begin{figure}[h]
	\centering
	\includegraphics[scale=0.1]{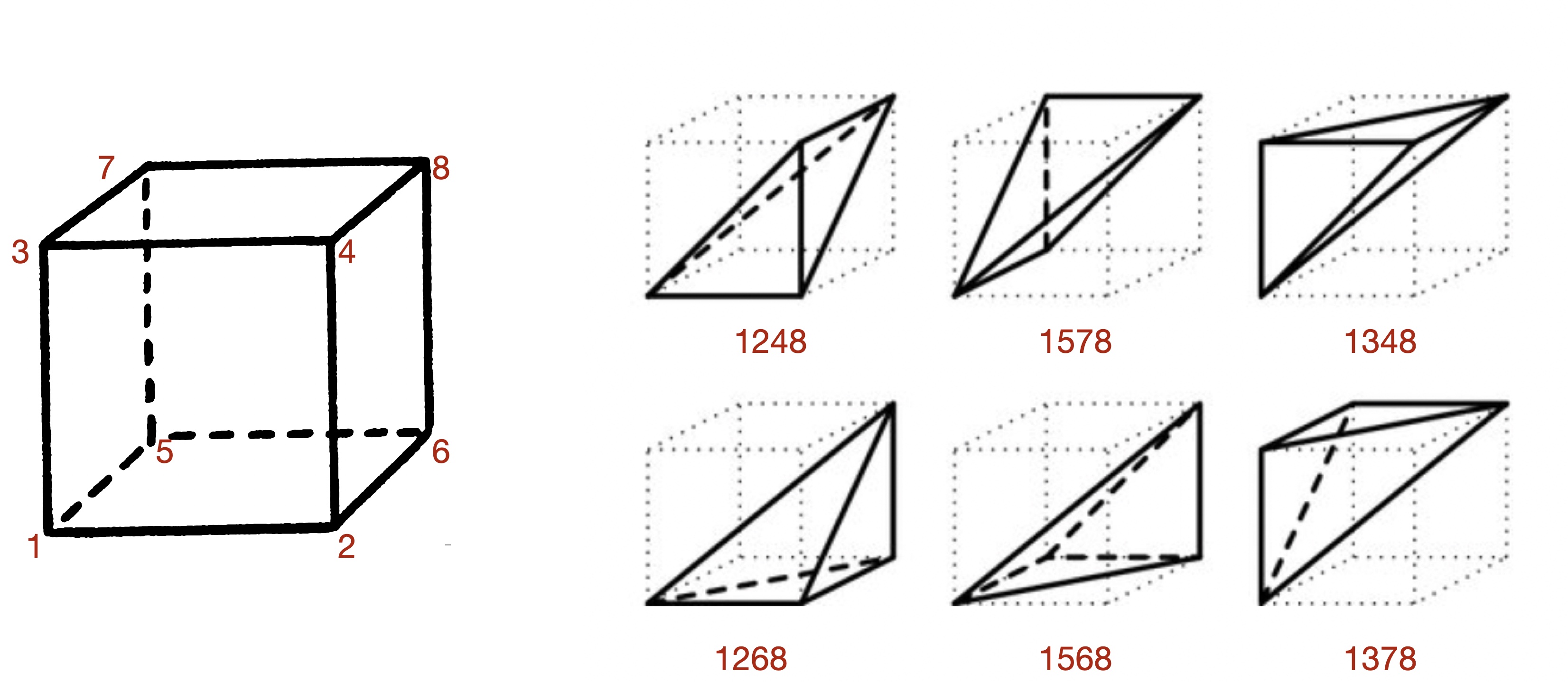} 
	\caption{Triangulation of a 3-d cube into six tetrahedra.  In this way, the tetrahedra all share one of the main diagonals of the cube. }\label{triangulation3D}
\end{figure}

To clarify how to triangulate a 4D hypercube into twenty-four 4-simplices, we begin with the eight 3D cubes, each of which can be subdivided into six irregular tetrahedra. This subdivision is achieved by first triangulating each face of the cube, and then selecting a vertex and connecting each triangle to that vertex. This process ensures that all resulting tetrahedra share one of the main diagonals of the cube, as illustrated in Figure \ref{triangulation3D}.
Subsequently, each of the 8 cubes is subdivided into six tetrahedra using the same approach. These distinct tetrahedra will then form 4-simplices by adding either vertex 1 or vertex 16. The set of vertices for the corresponding 4-simplices is:
\be
 \begin{aligned}
 \{v_1,\cdots,v_{24}\}=	&\{ (1,2,4,8,16),\,(1,2,4,12,16),\,(1,2,6,8,16),\,(1,2,6,14,16), \\
 	& (1,3,4,8,16),\,(1,3,4,12,16),\,(1,3,7,8,16),\,(1,3,7,15,16), \\
 	& (1,5,6,8,16),\,(1,5,6,14,16),\,(1,5,7,8,16),\,(1,5,7,15,16), \\
 	& (1,2,10,12,16),\,(1,2,10,14,16),\,(1,3,11,12,16),\,(1,3,11,15,16), \\
 	& (1,9,10,12,16),\,(1,9,10,14,16),\,(1,9,11,12,16),\,(1,9,11,15,16), \\
 	& (1,5,13,14,16),\,(1,5,13,15,16),\,(1,9,13,14,16),\,(1,9,13,15,16).\}
 \end{aligned}\label{v1to24}
\ee
In this way, all the 4-simplices share the main diagonal $(1,16)$ of the 4D hypercube. The dual graph representing this hypercube triangulation is shown in Figure \ref{hypercubedual}(a), where each vertex corresponds to a 4-simplex, each edge corresponds to a tetrahedron, and closed loops represent bulk faces. Boundary tetrahedra are distinguished with black segments, while segments of a different color represent bulk tetrahedra.


The periodic boundary condition is imposed by matching corresponding faces along spatial directions: 
\be
 \begin{aligned}
x-\text{direction cubes:}\quad 	& 1 \leftrightarrow 5,\,2 \leftrightarrow 6,\,3 \leftrightarrow 7,\,4 \leftrightarrow 8,\, 9 \leftrightarrow 13, \, 10 \leftrightarrow 14, \,11 \leftrightarrow 15,12 \leftrightarrow 16,\\
y-\text{direction cubes:}\quad 	& 1 \leftrightarrow 3,\,2 \leftrightarrow 4,\,5 \leftrightarrow 7,\,6 \leftrightarrow 8,\,9 \leftrightarrow 11,\,10 \leftrightarrow 12,\,13 \leftrightarrow 15,\,14 \leftrightarrow 16, \\
z-\text{direction cubes:}\quad & 1 \leftrightarrow 2,\,3 \leftrightarrow 4,\,5 \leftrightarrow 6,7 \leftrightarrow 8,\,9 \leftrightarrow 10,\,11 \leftrightarrow 12,\,13 \leftrightarrow 14,\,15 \leftrightarrow 16. 
 \end{aligned}
\ee
which is also illustrated in FIG. \ref{ratio}(b). With these identifications, some of the boundary tetrahedra in a single hypercube will become bulk tetrahedra. Specifically, the following pairs of tetrahedra are identified along the spatial directions: 
\be
\begin{aligned}
    (1,2,4,12)\leftrightarrow (5,6,8,16),\quad (1,3,4,12)\leftrightarrow (5,7,8,16),\quad (1,2,10,12)\leftrightarrow (5,6,14,16),\\
(1, 9, 10, 12)\leftrightarrow(5, 13, 14, 16),\quad (1, 3, 11, 12)\leftrightarrow (5, 7, 15, 16),\quad (1, 9, 11, 12)\leftrightarrow(5, 13, 15, 16),\\
(1, 2, 6, 14)\leftrightarrow (3, 4, 8, 16),\quad (1, 5, 6, 14)\leftrightarrow (3, 7, 8, 16), \quad (1, 2, 10, 14)\leftrightarrow (3, 4, 12, 16),\\
(1, 9, 10, 14)\leftrightarrow(3, 11, 12, 16),\quad (1, 5, 13, 14)\leftrightarrow(3, 7, 15, 16),\quad (1, 9, 13, 14)\leftrightarrow (3, 11, 15, 16),\\ 
(1, 3, 7, 15)\leftrightarrow(2, 4, 8, 16),\quad (1, 5, 7, 15)\leftrightarrow (2, 6, 8, 16),\quad(1, 
  3, 11, 15)\leftrightarrow (2, 4, 12, 16),\\
(1, 9, 11, 15)\leftrightarrow (2, 10, 12, 16),\quad (1, 5, 13, 15)\leftrightarrow(2, 6, 14, 
  16),\quad (1, 9, 13, 15)\leftrightarrow (2, 10, 14, 16).
\end{aligned}
\ee 
As a result of the periodic boundary condition made along the spatial directions, the dual diagram representing the hypercube triangulation transforms into Figure \ref{hypercubedual}(b).

\section{Regge action coupled with the scalar field} \label{ReggeWithAL}
The total action, denoted as $S_{\rm tot}$ in (\ref{Stot}), combines the spinfoam action and the scalar field in a manner analogous to the sums of the Regge action and the scalar field. The continuous expression for the gravity action on a psudo-Riemannian manifold is given by 
\be 
S_{\rm GR} =\frac{1}{2}\int \mathrm{~d}^4x R \sqrt{-g} = \frac{1}{2}\int R\mathrm{~d}V, \label{SWald}
\ee
Here, $R$ represents the scalar curvature, and $g$ is the determinant of the metric tensor. The derivation below is a modification of the derivation in \cite{PhysRevD.12.385} for the Lorentzian signature $(+,-,-,-)$ to the signature $(-,+,+,+)$ used in this paper. 

Initially, we consider the bulk face lying in the $t$-$z$ plane, defined as $=\{(txyz)|x=y=0\}$, and introduce the deficit angle $\delta$. When $\delta=0$, the coordinates $x$ and $y$ can be replaced by $r$ and $\phi$, resulting in the metric tensor:
\be 
g_{tt}=-1,\quad g_{zz}=1,\quad g_{rr}=1,\quad g_{\phi\phi} = r^2 = e^{2\lambda(r)}. \label{r0}
\ee
where all other components are zero. As in \cite{PhysRevD.12.385}, the defect is introduced by removing the "wedge" $2\pi-\delta\leq\phi<2\pi$ from the spacetime and smoothly extending $\phi$ to cover the remainder smoothly.  The metric is then given by:
\be 
g_{tt}=-1,\quad g_{zz}=1,\quad g_{rr}=1,\quad g_{\phi\phi} = \left(1-\frac{\delta}{2\pi}\right)^2r^2 = e^{2\lambda(r)}. \label{rneq0}
\ee 
The scalar curvature and the Riemann curvature tensor are defined by:
\be
\begin{gathered}
    R \equiv g^{\mu \nu} g^{\alpha \beta} R_{\mu \alpha \nu \beta} = g^{\mu\nu}R_{\mu\alpha\nu}{}^{\alpha},\\
    R_{\mu\nu\sigma}{ }^\rho = \Gamma^{\rho}{}_{\mu\sigma,\nu}-  \Gamma^{\rho}{}_{\nu\sigma,\mu} +  \Gamma^{\lambda}{}_{\sigma\mu}\Gamma^{\rho}{}_{\nu\lambda}-\Gamma^{\lambda}{}_{\sigma\nu}\Gamma^{\rho}{}_{\mu\lambda}.
\end{gathered}
\ee
One can obtain that
\be 
\begin{aligned}
R =-2[(\lambda')^2+\lambda''],\quad (-g)^{1 / 2} = e^\lambda, \quad \lambda'=\frac{\mathrm{d}\lambda}{\mathrm{d}r}
\end{aligned}
\ee  
from which the action in (\ref{SWald}) can be computed by
\be 
\frac{1}{2}\int R (-g)^{1/2} \mathrm{~d}r\mathrm{d}\phi = -2\pi \int^\infty_0 (e^\lambda)'' dr = \delta.
\ee 
with the definitions of $e^\lambda$ in (\ref{r0}) and (\ref{rneq0}) for $r=0$ and $r\neq 0$. Then
\be
S_{\rm GR} = \frac{1}{2} \int R(-g)^{1 / 2} d r d \phi d z d t & =\delta \iint d z d t = \mathrm{Ar} \delta,
\ee 
where $\mathrm{Ar}$ is the area of the face. Given any triangulation, the Regge action sums over the defects corresponding to the internal triangles $h$: 
\be
S_{\rm Regge}=\sum_h\mathrm{Ar}_h\delta_h.
\ee

Moveover, the continuum gravity action with the scalar field is given by \cite{Wald:1984rg}:
\be 
S_{\rm GR} + S_{\mathrm{KG}}=\frac{1}{2}\int \mathrm{d}^4x R\sqrt{-g} - \frac{1}{2} \int \sqrt{-g}\left(g^{a b} \nabla_a \phi \nabla_b \phi\right). 
\ee 
On a simplicial complex, we use the discretization of the scalar field in (\ref{AL}) suggested by \cite{REN1988661} (see Appendix \ref{Lg}). The discretized action of gravity with the scalar field is given by 
\be 
S_{\rm tot}=S_{\rm Regge} + S_{\rm L} = \sum_h\mathrm{Ar}_h\delta_h- \frac{1}{2} \sum_{b_{S S^{\prime}}} \rho_{S S^{\prime}}\left(\varphi_S-\varphi_{S^{\prime}}\right)^2. \label{Sall}
\ee 
The action of spinfoam with the scalar field in (\ref{Stot}) is analogous to this.

\section{Details about $\rho_{vv'}$}\label{Lg}

The continuum action of the scalar field is 
\be 
S_{\mathrm{KG}}=-\frac{1}{2} \int \sqrt{-g}\left(g^{\mu\nu} \nabla_\mu \varphi \nabla_\nu \varphi\right), \label{KGc}
\ee
and we use the following discretization of the scalar field suggested by \cite{REN1988661} 
\be 
S_{\mathrm{sc}} = - \frac{1}{2} \sum_{b_{v v^{\prime}}} \rho_{v v^{\prime}}\left(\varphi_v-\varphi_{v^{\prime}}\right)^2. \label{discreteKG}
\ee 
However the discrete action in \cite{REN1988661} is for the Euclidean signature, some suitable modification of $\rho_{v v^{\prime}}$ is needed for the Lorentzian signature. 

Let us compare $S_{\mathrm{sc}}$ to $S_{\rm KG}$ in the continuum approximation, so we can determine $\rho_{vv'}$. Given 4-simplices $v$ and $v'$ with $\varphi_v$ and $\varphi_{v'}$,
\be 
\varphi_v - \varphi_{v'} \simeq b_{vv'}^\mu \langle \partial_\mu\varphi\rangle,
\ee 
where $b^\mu_{vv'}$ is the displacement vector associate to (the centers of) $v,v'$, and $\langle \partial_\mu\varphi\rangle$ is understood as the mean value of the continuous $\partial_\mu\varphi$ in a neighborhood $U_{vv'}\subset v\cup v'$. The scalar field in (\ref{discreteKG}) can be rewritten as
\be 
S_{\mathrm{sc}} \simeq - \frac{1}{2} \sum_{b_{v v^{\prime}}} \rho_{v v^{\prime}}  b_{vv'}^\mu b_{vv'}^\nu \langle \partial_\mu\varphi \rangle\langle \partial_\nu\varphi\rangle. \label{KGd}
\ee 
By comparing the continuum action in (\ref{KGc}) and the discretized action in (\ref{KGd}), we propose the discretization
\be 
\int_{U_{vv'}} \mathrm{d}^4x \sqrt{-g} g^{\mu\nu}\partial_\mu\varphi\partial_\nu\varphi\simeq \rho_{v v^{\prime}}\left(\varphi_v-\varphi_{v^{\prime}}\right)^2\simeq   \rho_{v v^{\prime}}  b_{vv'}^\mu b_{vv'}^\nu\langle \partial_\mu\varphi \rangle\langle \partial_\nu\varphi\rangle.
\ee
For $\partial_\mu\varphi\simeq\langle\partial_\mu\varphi\rangle$ and $g_{\mu\nu}$ appriximately constant, we obtain
\be
\O_{U_{vv'}}\sqrt{-g} g^{\mu\nu}\simeq \rho_{v v^{\prime}}  b_{vv'}^\mu b_{vv'}^\nu,
\ee
where $\O_{U_{vv'}}$ is the coordinate volume of $U_{vv'}$. Moreover,
\be
\O_{U_{vv'}}\sqrt{-g} g^{\mu\nu} (b_{vv'})_\mu (b_{vv'})_\nu= \O_{U_{vv'}}\sqrt{-g}\, \lVert (b_{vv'})\rVert ^2 \simeq  
\rho_{vv^{\prime}} \lVert b_{vv'} \rVert^4 . \label{G7}
\ee 
For the timelike vector $\lVert b_{vv'}\rVert^2<0$ or spacelike vector $\lVert b_{vv'}\rVert^2 >0$, $\lVert b_{vv'} \rVert^4>0$ always holds. Therefore, we can rewrite (\ref{G7}) as 
\be 
\O_{U_{vv'}} \sqrt{-g}   \simeq  \rho_{v v^{\prime}} \lVert b_{vv'} \rVert^2 
\ee 
We approximate $\O_{U_{vv'}} \sqrt{-g} $ by $|b_{vv'}||V_{vv'}|$, where $|b_{vv'}|=\sqrt{ |\lVert b_{vv'} \rVert^2|}$ and $|V_{vv'}|$ is the 3-volume of the tetrahedron shared by $vv'$. More specifically, we use $|b_{vv'}|=|b_{ve}|+|b_{v'e}|$ where $|b_{ve}|>0$ is the distance from the centroid of the 4-simplex $v$ to the centroid of the tetrahedron $e\subset \partial v$. So up to an overall constant, $|b_{vv'}||V_{vv'}|$ is the sum of two volumes in $v$ and $v'$ associated to $e$. As a result, we obtain
\be
\rho_{vv'}=\sgn\lt(b_{vv'} \rt)\frac{|V_{vv'}|}{|b_{vv'}|},\qquad \sgn \left(b_{vv^{\prime}}\right) = \left\{\begin{matrix}
    -1, \quad b_{vv'}\text{ is timelike}\\
     +1, \quad b_{vv'}\text{ is spacelike}
\end{matrix}\right. 
\ee
In this work, the flat geometries on the hypercube and double-hypercube complexes only give spacelike $b_{vv'}$, so $\sgn \left(b_{vv^{\prime}}\right)=+1$ in the numerical computation. 

Below we give two examples of $\rho_{vv'}$'s explicit expressions. In the hypercube complex, $\rho_{vv'}$ can be expressed in terms of $(a_f,a_i, h)$. The 4-simplices $v=(1, 2, 4, 8, 16)$ and $v'=(1, 2, 4, 12, 16)$ share a tetrahedron $e=(1,2,4,16)$, the corresponding $\rho_{vv'}$ is 
\be
\rho_{vv'}=\frac{10 a_i^2 \sqrt{(a_i+a_f)^2-4 h^2}}{3 \left(\sqrt{83 a_i^2-30 a_i a_f+3 a_f^2-4 h^2}+\sqrt{11 a_i^2-30 a_i a_f+43 a_f^2-36 h^2}\right)}.
\ee 
In the double-hypercube complex, $\rho_{vv'}$ can be expressed in terms of $(a_i, a_m, a_f, h_i, h_f)$. The 4-simplices $v=(1, 17, 18, 20, 24)$ and $v'=(1, 5, 21, 22, 24)$ share a tetrahedron $e=(1, 17, 18, 20)$ identified with $e=(5, 21, 22, 24)$, the corresponding $\rho_{vv'}$ is 
\be
\rho_{vv'}=\frac{10 a_f^2 \sqrt{4 h_f^2-(a_f - a_m)^2}}{3 \left(\sqrt{3 a_f^2+2 a_f(5 a_m+2)+a_m (43 a_m+20)-36 h_f^2+4}+\sqrt{83 a_f^2+30 a_f a_m+3 a_m^2-4 h_f^2}\right)}.
\ee

\end{document}